\documentclass[twocolumn,tighten,floatfix]{aastex631} 
\pdfoutput=1 
\usepackage{amsmath,amstext}

\usepackage{hyperref}
\usepackage{xcolor}
\usepackage{xspace}
\usepackage{hhline}
\usepackage[caption=false]{subfig}

\usepackage{datatool}
\DTLsetseparator{ = } 

\definecolor{mulberry}{HTML}{A93C93}

\hypersetup{
colorlinks=true,
linkcolor=mulberry,
urlcolor=mulberry,
citecolor=mulberry
}

\graphicspath{{figures/}}

\newcommand{\Dz}{\Delta z}
\newcommand{\dz}{\Dz/(1+z)}
\newcommand{\zgaia}{z_\mathrm{Gaia}}
\newcommand{\zquaia}{z_\mathrm{Quaia}}
\newcommand{\zknn}{z_{k\mathrm{NN}}}
\newcommand{\zsdss}{z_\mathrm{SDSS}}

\newcommand{\knn}{$k$NN\xspace}
\newcommand{\Gaia}{\textsl{Gaia}\xspace}
\newcommand{\Gaiapurer}{\Gaia DR3 {`Purer'}\xspace}
\newcommand{\unWISE}{\textsl{unWISE}\xspace}
\newcommand{\SDSS}{\textsl{SDSS}\xspace}
\newcommand{\Catalog}{\Gaia--\unWISE Quasar Catalog\xspace}
\newcommand{\catalog}{\Catalog}
\newcommand{\cat}{Quaia\xspace}

\newcommand{\Mpch}{h^{-1}\,\text{Mpc}}

\newcommand{\new}[1]{#1}

\newcommand\minput[1]{%
  \input{#1}%
  \ifhmode\ifnum\lastnodetype=11 \unskip\fi\fi}

\sloppypar
\begin{document}

\begin{filecontents*}{quantities.txt}
N_gall = 6,649,162
N_gall_wqsoc = 6,375,063
N_gall_flagsqsoc0 = 302,297
N_gall_flagsqsoc0or16 = 1,302,677
N_gpurer = 1,942,825
N_gpurer_wqsoc = 1,729,625
N_gsup = 1,518,782
N_gclean = 1,414,385
p_cut_gsup_gclean = 7
N_gcatlo = 755,850
N_gcathi = 1,295,502
N_gcathi_flagsqsoc0 = 276,293
p_gcathi_flagsqsoc0 = 21
N_gcathi_flagsqsoc0or16 = 507,578
p_gcathi_flagsqsoc0or16 = 39
N_squasars_unwise = 343,074
N_squasars_sup = 246,122
N_sqall = 638,083
N_sqclean = 243,206
N_eboss = 409,341
N_sgals_unwise = 600,897
N_sgals_sup = 1,316
N_stars_unwise = 482,080
N_sstars_sup = 2,276
N_mcs_unwise = 11,770
N_mcs_sup = 9,927
z_med_gall = 1.67
z_med_gcatlo = 1.45
z_med_gcathi = 1.47
zintermediate = 2.5
N_above_zintermediate_gcatlo = 77,337
p_above_zintermediate_gcatlo = 10
N_above_zintermediate_gcathi = 132,417
p_above_zintermediate_gcathi = 10
smad_dz_Ghi = 0.008
smad_Dz_Ghi = 0.023
Gbright = 19.0
dzlo = 0.01
p_outliers_zspz_dzlo_Gbright = 12
p_acc_zspz_dzlo_Gbright = 88
p_outliers_zgaia_dzlo_Gbright = 10
p_acc_zgaia_dzlo_Gbright = 90
dzmid = 0.1
p_outliers_zspz_dzmid_Gbright = 4
p_acc_zspz_dzmid_Gbright = 96
p_outliers_zgaia_dzmid_Gbright = 7
p_acc_zgaia_dzmid_Gbright = 93
dzhi = 0.2
p_outliers_zspz_dzhi_Gbright = 3
p_acc_zspz_dzhi_Gbright = 97
p_outliers_zgaia_dzhi_Gbright = 7
p_acc_zgaia_dzhi_Gbright = 93
Glo = 20.0
p_outliers_zspz_dzlo_Glo = 25
p_acc_zspz_dzlo_Glo = 75
p_outliers_zgaia_dzlo_Glo = 25
p_acc_zgaia_dzlo_Glo = 75
p_outliers_zspz_dzmid_Glo = 10
p_acc_zspz_dzmid_Glo = 90
p_outliers_zgaia_dzmid_Glo = 19
p_acc_zgaia_dzmid_Glo = 81
p_outliers_zspz_dzhi_Glo = 6
p_acc_zspz_dzhi_Glo = 94
p_outliers_zgaia_dzhi_Glo = 18
p_acc_zgaia_dzhi_Glo = 82
Ghi = 20.5
p_outliers_zspz_dzlo_Ghi = 37
p_acc_zspz_dzlo_Ghi = 63
p_outliers_zgaia_dzlo_Ghi = 38
p_acc_zgaia_dzlo_Ghi = 62
p_outliers_zspz_dzmid_Ghi = 16
p_acc_zspz_dzmid_Ghi = 84
p_outliers_zgaia_dzmid_Ghi = 30
p_acc_zgaia_dzmid_Ghi = 70
p_outliers_zspz_dzhi_Ghi = 9
p_acc_zspz_dzhi_Ghi = 91
p_outliers_zgaia_dzhi_Ghi = 28
p_acc_zgaia_dzhi_Ghi = 72
p_outliers_gall_zgaia_dzlo_Gbright = 9
p_acc_gall_zgaia_dzlo_Gbright = 91
p_outliers_gall_zgaia_dzmid_Gbright = 7
p_acc_gall_zgaia_dzmid_Gbright = 93
p_outliers_gall_zgaia_dzhi_Gbright = 6
p_acc_gall_zgaia_dzhi_Gbright = 94
p_outliers_gall_zgaia_dzlo_Glo = 25
p_acc_gall_zgaia_dzlo_Glo = 75
p_outliers_gall_zgaia_dzmid_Glo = 19
p_acc_gall_zgaia_dzmid_Glo = 81
p_outliers_gall_zgaia_dzhi_Glo = 18
p_acc_gall_zgaia_dzhi_Glo = 82
p_outliers_gall_zgaia_dzlo_Ghi = 38
p_acc_gall_zgaia_dzlo_Ghi = 62
p_outliers_gall_zgaia_dzmid_Ghi = 30
p_acc_gall_zgaia_dzmid_Ghi = 70
p_outliers_gall_zgaia_dzhi_Ghi = 28
p_acc_gall_zgaia_dzhi_Ghi = 72
factor_reduction_outliers_dzlo_Glo = ${\sim}1\times$
factor_reduction_outliers_dzmid_Glo = ${\sim}2\times$
factor_reduction_outliers_dzhi_Glo = ${\sim}3\times$
factor_reduction_outliers_dzlo_Ghi = ${\sim}1\times$
factor_reduction_outliers_dzmid_Ghi = ${\sim}2\times$
factor_reduction_outliers_dzhi_Ghi = ${\sim}3\times$
p_zreliable_gall = 20
p_acc_gall_zgaia_dzlo = 53
p_acc_gall_zreliable_zgaia_dzlo = 86
p_acc_gall_zgaia_dzmid = 62
p_acc_gall_zreliable_zgaia_dzmid = 91
p_acc_gall_zgaia_dzhi = 65
p_acc_gall_zreliable_zgaia_dzhi = 92
N_removed_pmcut = 39470
p_removed_pmcut = 2.6
factor_reduction_contaminants = ${\sim}4\times$
p_sqall_excluded_clean = 1.2
Avhi = 0.5
area_below_Avhi = 30277.52 deg$^2$
fsky_below_Avhi = 0.73
volume_effective_gcatlo_below_Avhi = 3.19 $(h^{-1}\,\text{Gpc})^3$
volume_effective_gcathi_below_Avhi = 7.67 $(h^{-1}\,\text{Gpc})^3$
area_sdss = 4808 deg$^2$
fsky_sdss = 0.12
volume_effective_sdss = 5.72 $(h^{-1}\,Gpc)^3$
factor_area_belowAvhi_sdss = ${\sim}6.5\times$
factor_volume_effective_gcathi_sdss = ${\sim}1.3\times$
area_eboss = 4808 deg$^2$
fsky_eboss = 0.12
volume_effective_eboss = 3.41 $(h^{-1}\,Gpc)^3$
factor_area_belowAvhi_eboss = ${\sim}6.5\times$
factor_volume_effective_gcathi_eboss = ${\sim}1.3\times$
rmse_fractional_residuals_Glo = 0.49
rmse_fractional_residuals_Ghi = 0.44
color_cut_str = (G-W1) &> 2.15 \\ (W1-W2) &> 0.4 \\ (BP-G) &> -0.3 \\ (G-W1) + 1.2\,(W1-W2) &> 3.4
p_seyferts_Mi-22 = 8
\end{filecontents*}
\DTLloaddb[noheader, keys={thekey,thevalue}]{quantities}{quantities.txt}
\DTLnewdbonloadfalse 
\begin{filecontents*}{quantities_comparison.txt}
p_reliable_wiseps = 59
Avhi = 0.5
N_gcathi = 1,234,715
area_gcathi = 30275.84 deg$^2$
fsky_gcathi = 0.73
nbar_gcathi = 40.78 deg$^-2$
volume_effective_gcathi = 7.08 $(h^{-1}\,Gpc)^3$
volume_spanning_gcathi = 143.78 $(h^{-1}\,Gpc)^3$
N_gpurer = 1,286,788
area_gpurer = 30270.80 deg$^2$
fsky_gpurer = 0.73
nbar_gpurer = 42.51 deg$^-2$
volume_effective_gpurer = 6.50 $(h^{-1}\,Gpc)^3$
volume_spanning_gpurer = 143.76 $(h^{-1}\,Gpc)^3$
N_wiseps = 1,130,925
area_wiseps = 22964.75 deg$^2$
fsky_wiseps = 0.56
nbar_wiseps = 49.25 deg$^-2$
volume_effective_wiseps = 7.32 $(h^{-1}\,Gpc)^3$
volume_spanning_wiseps = 109.06 $(h^{-1}\,Gpc)^3$
N_sdss = 297,940
area_sdss = 10575.94 deg$^2$
fsky_sdss = 0.26
nbar_sdss = 28.17 deg$^-2$
volume_effective_sdss = 1.18 $(h^{-1}\,Gpc)^3$
volume_spanning_sdss = 50.23 $(h^{-1}\,Gpc)^3$
N_eboss = 190,263
area_eboss = 5603.12 deg$^2$
fsky_eboss = 0.14
nbar_eboss = 33.96 deg$^-2$
volume_effective_eboss = 1.01 $(h^{-1}\,Gpc)^3$
volume_spanning_eboss = 26.61 $(h^{-1}\,Gpc)^3$
zmed_gcathi = 1.48
frac_acc_dzlo_gcathi = 0.63
frac_acc_dzmid_gcathi = 0.84
zmed_gpurer = 1.61
frac_acc_dzlo_gpurer = 0.62
frac_acc_dzmid_gpurer = 0.70
zmed_wiseps = 1.41
frac_acc_dzlo_wiseps = 0.12
frac_acc_dzmid_wiseps = 0.76
zmed_sdss = 1.67
zmed_eboss = 1.49
p_nuv_galex = 32
p_nuv_fuv_galex = 16
p_panstarrs = 75
\end{filecontents*}
\DTLloaddb[noheader, keys={thekey,thevalue}]{quantities}{quantities_comparison.txt}

\newcommand{\val}[1]{%
    \DTLgetvalueforkey{\scratchmacro}{thevalue}{quantities}{thekey}{#1}%
    \DTLifnull{\scratchmacro}{UUU}{\scratchmacro}\xspace
}

\newcommand{\Ghi}{\val{Ghi}}
\newcommand{\Glo}{\val{Glo}}
\newcommand{\Gmax}{20.6}
\newcommand{\colorcutstr}{\val{color_cut_str}}

\shorttitle{\cat: the \Catalog}

\title{\cat, the \Catalog: An All-Sky Spectroscopic Quasar Sample}

\author[0000-0001-8764-7103]{Kate Storey-Fisher}
\affiliation{Center for Cosmology and Particle Physics, Department of Physics, New York University, 726 Broadway, New York, NY 10003, USA}

\author[0000-0003-2866-9403]{David W. Hogg}
\affiliation{Center for Cosmology and Particle Physics, Department of Physics, New York University, 726 Broadway, New York, NY 10003, USA}
\affiliation{Center for Computational Astrophysics, Flatiron Institute, 162 Fifth Avenue, New York, NY, 10010, USA}
\affiliation{Max Planck Institute for Astronomy, K{\"o}nigstuhl 17, D-69117 Heidelberg, Germany}

\author[0000-0003-4996-9069]{Hans-Walter Rix}
\affiliation{Max Planck Institute for Astronomy, K{\"o}nigstuhl 17, D-69117 Heidelberg, Germany}

\author[0000-0003-2895-6218]{Anna-Christina Eilers}
\affiliation{MIT Kavli Institute for Astrophysics and Space Research, 77 Massachusetts Avenue, Cambridge, 02139, Massachusetts, USA}

\author[0000-0002-3255-4695]{Giulio Fabbian}
\affiliation{Center for Computational Astrophysics, Flatiron Institute, 162 Fifth Avenue, New York, NY, 10010, USA}
\affiliation{School of Physics and Astronomy, Cardiff University, The Parade, Cardiff, Wales CF24 3AA, United Kingdom}

\author[0000-0003-1641-6222]{Michael R. Blanton}
\affiliation{Center for Cosmology and Particle Physics, Department of Physics, New York University, 726 Broadway, New York, NY 10003, USA}

\author[0000-0002-4598-9719]{David Alonso}
\affiliation{Department of Physics, University of Oxford, Denys Wilkinson Building, Keble Road, Oxford OX1 3RH, United Kingdom}

\correspondingauthor{Kate Storey-Fisher}
\email{k.sf@nyu.edu}

\begin{abstract}
We present a new, all-sky quasar catalog, \cat, that samples the largest comoving volume of any existing spectroscopic quasar sample.
The catalog draws on the \val{N_gall} quasar candidates identified by the \Gaia mission that have redshift estimates from the space observatory's low-resolution blue photometer/red photometer spectra.
This initial sample is highly homogeneous and complete, but has low purity, and $\val{p_outliers_zgaia_dzhi_Glo}\%$ of even the bright ($G<\Glo$) confirmed quasars have discrepant redshift estimates ($|\dz| > \val{dzhi}$) compared to those from the Sloan Digital Sky Survey (\SDSS).
In this work, we combine the \Gaia candidates with \unWISE infrared data (based on the Wide-field Infrared Survey Explorer survey) to construct a catalog useful for cosmological and astrophysical quasar studies.
We apply cuts based on proper motions and colors, reducing the number of contaminants by \val{factor_reduction_contaminants}.
We improve the redshifts by training a $k$-Nearest Neighbors model on \SDSS redshifts, and achieve estimates on the $G<\Glo$ sample with only $\val{p_outliers_zspz_dzhi_Glo}\%$ ($\val{p_outliers_zspz_dzmid_Glo}\%$) catastrophic errors with $|\dz| > \val{dzhi}$ ($\val{dzmid}$), a reduction of \val{factor_reduction_outliers_dzhi_Glo} (\val{factor_reduction_outliers_dzmid_Glo}) compared to the \Gaia redshifts.
The final catalog has \val{N_gcathi} quasars with $G<\Ghi$, and \val{N_gcatlo} candidates in an even cleaner $G<\Glo$ sample, with accompanying rigorous selection function models.
We compare \cat to existing quasar catalogs, showing that its large effective volume makes it a highly competitive sample for cosmological large-scale structure analyses.
The catalog is publicly available at \url{https://doi.org/10.5281/zenodo.10403370}. 
\end{abstract}

\section{Introduction}

Quasars are powerful tools for many fields of astrophysics. 
They are key probes of accretion physics (e.g. \citealt{SunyaevZeldovich1970, yu_quasar_2020}), which informs the evolution of active galactic nuclei (AGNs). 
The evolution of quasars and their host galaxies are intertwined, giving insight into supermassive black hole growth (e.g. \citealt{hopkins_unified_2006}) as well as massive galaxy formation (e.g. \citealt{kormendy_coevolution_2013}).
Studies of the quasar distribution can also be used to understand black hole evolution (e.g. \citealt{powell_clustering_2020}) and halo masses and environmental effects (e.g. \citealt{dipompeo_characteristic_2017}).
Quasars can also be utilized as background sources for cosmic phenomena such as gravitational lenses (e.g. \citealt{claeskens_gravitational_2002}), and quasar spectra encode the properties of the intergalactic medium via the Ly$\alpha$ forest (e.g. \citealt{rauch_lyman_1998}). 

Quasars are key tracers for large-scale structure cosmology.
They reside in peaks of the dark matter distribution and their clustering can be used to measure cosmological parameters, including the growth rate of structure $f\sigma_8$ (e.g. \citealt{garcia-garcia_growth_2021, alonso_constraining_2023}), the Hubble distance $D_H$ (e.g. \citealt{hou_completed_2020}), primordial non-Gaussianity (e.g. \citealt{leistedt_constraints_2014, castorina_redshift-weighted_2019, krolewski_constraining_2023}), and the baryon density $\Omega_b$ (e.g. \citealt{yahata_large-scale_2005}). 
Cross-correlations between quasars and other tracers provide measurements of key cosmological quantities, such as with photometric galaxy samples to measure the baryon acoustic feature (e.g. \citealt{ata_clustering_2018}), with cosmic microwave background (CMB) lensing to constrain quasar bias and the growth of structure (e.g. \citealt{sherwin_atacama_2012}), and with foreground galaxies as a probe of weak lensing (e.g. \citealt{menard_cosmological_2002, scranton_detection_2005, zarrouk_baryon_2021}).
They can also be used as standardizable candles to measure the expansion rate of the universe (e.g. \citealt{setti_hubble_1973, risaliti_hubble_2015, lusso_quasars_2020}).
Finally, given the large volume typically covered by quasar samples, the quasar distribution provides a test of the cosmological principle of isotropy and homogeneity (e.g. \citealt{secrest_test_2021, dam_testing_2022, quaia-homogeneity}). 

Many surveys have observed and cataloged quasars, with around 1 million spectroscopically identified and several million when including photometric samples. 
The Sloan Digital Sky Survey (\SDSS) Data Release 16 includes a highly complete catalog of 750,414 quasars with spectroscopic redshifts \citep{lyke_sloan_2020}.
Photometric surveys observe a much larger number of quasars, at the expense of low redshift accuracy; nearly 3 million quasars with reliable photometric redshifts have been cataloged \citep{kunsagi-mate_photometric_2022}, including with the Wide-field Infrared Survey Explorer (WISE; \citealt{wright_wide-field_2010}), which imaged the entire sky and Pan-STARRS, \citep{chambers_pan-starrs1_2019} which observed three-quarters of the sky.
\cite{shu_catalogues_2019} combined photometry from \Gaia DR2 and unWISE \citep{lang_unwise_2014} to identify 2.7 million AGN candidates and estimate their photometric redshifts.
Upcoming surveys will observe even more quasars: the Dark Energy Spectroscopic Instrument (DESI; \citealt{Aghamousa2016}) expects to obtain spectra for 3 million quasars, and the Rubin Observatory's LSST will photometrically observe upward of 10 million quasars \citep{ivezic_lsst_2016}.
However, none of these quasar catalogs is both all-sky and contains precise redshift information.
The recently released \Gaia DR3 quasar candidates \citep{gaia_collaboration_gaia_2023} constitute a new sample that promises to fill this gap. 

The \Gaia quasar sample presents a new opportunity to explore these science topics. 
While the \Gaia satellite was designed to map stars in the Milky Way \citep{gaia_collaboration_gaia_2016}, it broadly observes bright objects in the sky, which includes many extragalactic sources. 
Previous work identified a small number of quasars in earlier \Gaia data releases, including identification based solely on their astrometric properties \citep{heintz_unidentified_2018, heintz_spectroscopic_2020}.
In DR3, the \Gaia collaboration released a sample of 6,649,162 quasar candidates that were incidentally observed during the survey \citep{delchambre_gaia_2023, gaia_collaboration_gaia_2023, gaia_collaboration_gaia_2023-1}.
The sources cover the entire sky and have \Gaia blue photometer (BP)/red photometer (RP) spectra, low-resolution spectra covering the wavelength range of 330--1050 nm. 
These spectra allow for redshift estimates of the sources, with $\val{p_acc_gall_zreliable_zgaia_dzlo}\%$ having a precision of $|\dz| < \val{dzlo}$ compared to \SDSS redshifts when no processing issues affect the redshift estimation (\texttt{flags\_qsoc} = 0 or \texttt{flags\_qsoc} = 16), which is the case for $\val{p_zreliable_gall}\%$ of the sample; for the full sample including sources with redshift warning flags set, this percentage of high-precision redshifts decreases to $\val{p_acc_gall_zgaia_dzlo}\%$.
While not as precise as high-resolution spectroscopic redshifts, they are significantly better than photometric redshifts. 
The median redshift of the sample is $z=\val{z_med_gall}$. 
The \Gaia quasar candidate sample was constructed for completeness over purity, and has an estimated purity of 52\%; the \Gaia Collaboration also suggests criteria for a higher purity ($\sim$95\%) subcatalog of $\sim$1.9 million quasars.
Overall, the sample presents an unprecedented resource for quasar science and cosmology.

There are two main issues with this raw \Gaia sample.
First, the sample contains a large number of non-quasar contaminants.
Second, a significant fraction of the redshift estimates are catastrophic errors, due to emission line misidentification given the limitations of the low-resolution spectra.
Understanding and eliminating sample contaminants matters greatly in identifying the most extreme (e.g. brightest or most luminous) quasars, which has been addressed in the AllBRICQS catalog \citep{onken_allbricqs_2023} that draws on Gaia quasar candidates. 
In this work, we construct a clean quasar catalog across the full magnitude range with lower contamination and improved redshift estimates, with the particular goal of building a catalog appropriate for large-scale structure analyses as well as other quasar science.
For both of these, we rely on crossmatches with WISE observations of the quasars \citep{wright_wide-field_2010}, which adds key infrared (IR) information.
To filter out contaminants, we apply color cuts based on the \Gaia and WISE photometry, as well as a proper motion cut.
To improve the redshifts, we identify quasars that are also observed by \SDSS, for which we have highly precise spectroscopic redshifts, and train a $k$-Nearest-Neighbors ($k$NN) model based on their photometry and \Gaia redshift estimates.
Further, the \Gaia quasar candidate sample has strong systematic imprints from various observational effects, such as Galactic dust.
\new{To model these systematics so that their effects can be mitigated in analyses of the catalog,} we fit a model for the selection function based on observational templates using a Gaussian process. \new{We release both the catalog and selection function as publicly accessible data products.}

This paper is organized as follows.
In \S\ref{sec:data}, we describe the initial data sets used in the construction of the catalog.
The construction of the catalog is detailed in \S\ref{sec:construction}.
In \S\ref{sec:catalog}, we present the final catalog and perform verification and comparisons to other samples, and outline the data format.
We summarize the catalog and describe the access to the data in \S\ref{sec:summary}.

\section{Initial Data Sets}
\label{sec:data}

\begin{figure}
    \centering
    \includegraphics[width=0.45\textwidth]{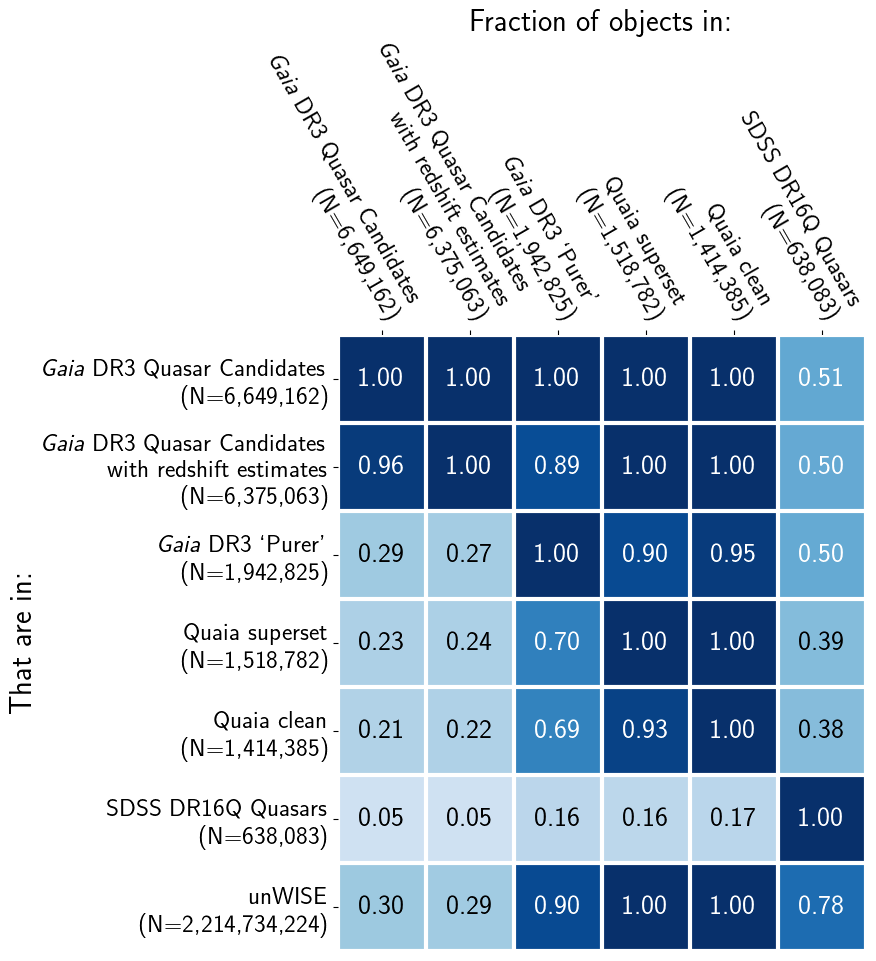}
    \caption{A summary of the overlaps between the various data sets and subsamples used in this work. The values describe the fraction of objects in each column's sample that are in each row's sample. Note that we only list \unWISE as a row because the inverse is not relevant to this work.}
    \label{fig:frac_matrix}
\end{figure}

\subsection{\Gaia DR3 quasar candidate sample}
\label{sec:data_gaia}

\begin{figure}
    \centering
    \includegraphics[width=0.45\textwidth]{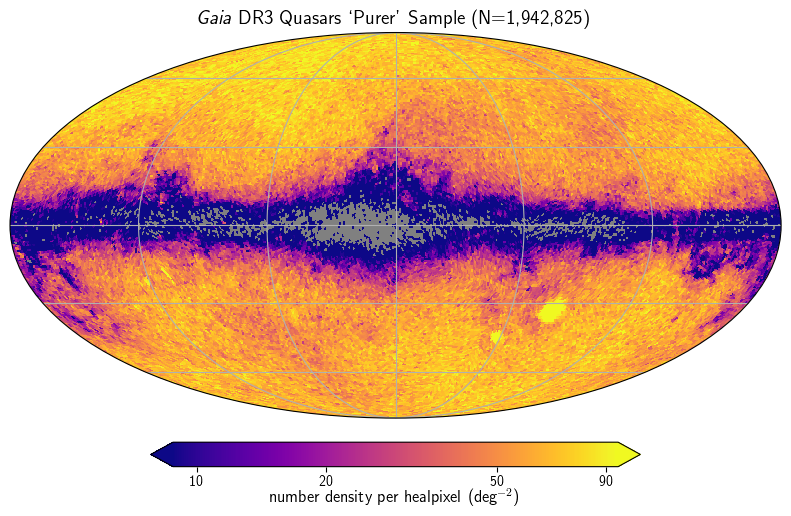}
    \caption{Sky distribution of the quasar candidates in the \Gaiapurer quasar sample, in Galactic coordinates and displayed using a Mollweide projection.}
    \label{fig:gaia_2d_purer}
\end{figure}

While performing its all-sky survey of the Milky Way, the \Gaia satellite \citep{gaia_collaboration_gaia_2016} also observed millions of extragalactic objects.
These sources---both quasar and galaxy candidates---were first released in \Gaia DR3 \citep{gaia_collaboration_gaia_2023, gaia_collaboration_gaia_2023-1}.
\Gaia obtained BP/RP spectra of the sources, which are low-resolution spectra with relatively narrow wavelength ranges; BP covers 330--680 nm and has $30 \leq R \leq 100$ and RP covers 640--1050 nm \citep{carrasco_internal_2021} with $70 \leq R \leq 100$.
The raw spectra are not released by \Gaia (besides a small subsample---the rest will be released in \Gaia DR4), but redshift estimates and other derived information are contained in the catalogs.

The quasar candidates were selected based on multiple classifiers and criteria, described in detail in \cite{gaia_collaboration_gaia_2023}.
The majority (5.5 million) of the quasar candidates were identified with the Discrete Source Classifier (DSC) module (detailed in \cite{delchambre_gaia_2023}, a machine-learning model that takes as input the source's BP/RP spectrum, $G$-band magnitude, $G$-band variability, parallax, and proper motion, and outputs a class label trained on \SDSS spectroscopic classifications.
Given these \SDSS labels, the results of this module will inherit many of the same selection effects as \SDSS.
DSC is estimated to have a completeness of over $90\%$ and a purity of around $g24\%$ for quasars.
Another machine learning model selected over 1 million sources based on their variability, as active nuclei have time-variable accretion; the model inputs were statistics of time series data in all \Gaia bands as well as photometric and astrometric quantities, as detailed in \cite{rimoldini_gaia_2023}. 
Additionally, a set of nearly 1 million sources was selected based on their surface brightness profile; this selection used existing major quasar catalogs to compile an initial list of sources, which were then processed by the \Gaia surface brightness profile module \citep{ducourant_gaia_2023}.
This module included quasars in the candidates catalog which passed certain criteria, including having \Gaia observations covering $>86\%$ of the source's surface area and a confident assessment (positive or negative) of host galaxy presence.
Finally, the 1.6 million sources used to define the \Gaia-CRF3 celestial reference frame were contributed, which are based on crossmatches of \Gaia to external quasar catalogs.
A large fraction of sources are identified as quasars by multiple of these methods; the overlapping contributions are shown in Figure 3 of \cite{gaia_collaboration_gaia_2023}.
The full quasar candidate sample contains \val{N_gall} sources\footnote{The \Gaia DR3 quasar candidates sample (and all other \Gaia data) can be downloaded at \url{https://gea.esac.esa.int/archive} with table name \texttt{gaiadr3.qso\_candidates}.}, selected for high completeness, but with a low purity estimated to be around 52\% \citep{gaia_collaboration_gaia_2023}.
We show the overlaps between this \Gaia quasar candidate sample and other samples and subsamples used and constructed in this work in Figure~\ref{fig:frac_matrix}.

Most of the quasar candidates (\val{N_gall_wqsoc}) are assigned redshifts using the Quasar Classifier (QSOC) module, which uses a chi-squared approach on the quasars' BP/RP spectra compared to composite spectra from \SDSS DR12Q \citep{delchambre_gaia_2023}.
We refer to these \Gaia redshift estimates as $\zgaia$.
Many of these redshifts are determined by a single line due to the narrow spectral range, resulting in aliasing issues when lines are misidentified (see Figure 15 in \citealt{delchambre_gaia_2023}).
An estimated 63.7\% of the redshifts have $|\Dz| < 0.1$, increasing to 97.6\% for quasar candidates with no redshift warning flags (this is the case for nearly $80\%$ of quasars with $G<18.5$, but decreases to less than 20\% for $G>19.5$).

\cite{gaia_collaboration_gaia_2023} provide a query to select a purer subsample of the quasar candidates.
It requires higher quasar probability thresholds from the various classifiers and excludes surface-brightness-selected galaxies that have close neighbors.
This results in \val{N_gpurer} sources with an estimated purity of 95\%; 1.7 million of these have Gaia redshifts. 
The sky distribution of this sample, which we call the \Gaiapurer sample, is shown in Figure~\ref{fig:gaia_2d_purer}.
The \Gaiapurer sample has a low density in the Galactic plane; \new{we speculate that} this is largely due to dust extinction making sources too faint to observe at low Galactic latitudes.
\Gaiapurer also has significant overdensities around the LMC and SMC, as the sample still contains stellar contaminants.

For our analysis, we start with the full quasar candidate sample, rather than the \Gaiapurer sample or cutting on other \Gaia pipeline flags, to allow greater completeness and minimize reproducing biases; we compare our catalog with the \cite{gaia_collaboration_gaia_2023} \Gaiapurer subsample in \S\ref{sec:comparison}.
We construct a \emph{superset} of our catalog (which is a subset of the \Gaia quasar candidates sample) that contains all the information needed for catalog construction: we require that sources are in the \Gaia quasar candidates table, have \Gaia $G$, $BP$, and $RP$ measurements, \unWISE $W1$ and $W2$ observations (described in \S\ref{sec:data_wise}), \Gaia-estimated QSOC redshifts, and a maximum $G$ magnitude of $G < \Gmax$.
This magnitude cut was chosen to be slightly deeper than our desired catalog magnitude limit of $G<\Ghi$, in order to provide a buffer for redshift estimation.
This results in a superset with \val{N_gsup} sources.
We call our final catalog \cat, so we refer to this as the \cat superset.

\subsection{\unWISE Quasar Sample}
\label{sec:data_wise}

We use the \unWISE reprocessing \citep{lang_unwise_2014,meisner_unwise_2019} of WISE \citep{wright_wide-field_2010} to contribute IR photometry to \Gaia sources.
The \unWISE coadds combine data from NEOWISE \citep{mainzer_preliminary_2011} with the original WISE survey, providing a time baseline 15 times longer.
Compared to the original \textsl{AllWISE} catalog, \unWISE has deeper imaging and improved modeling of crowded fields.
The \unWISE catalog \citep{schlafly_unwise_2019} contains measurements in the $W1$ (3.4 $\mu$m) and  $W2$ (4.6 $\mu$m) bands for over 2 billion sources.
We do not use the $W3$ and $W4$ bands as these do not go as deep as we need.
We perform a crossmatch of the \Gaia quasar candidate sample to \unWISE sources within 1\arcsec\footnote{We use NOIRLab's crossmatch service to perform this operation, available at \url{https://datalab.noirlab.edu/xmatch.php}.}.
We also crossmatch the \SDSS training and validation samples (\S\ref{sec:data_sdss_quasars}, \S\ref{sec:data_contaminants}) to \unWISE.

When combined with optical photometry, \unWISE IR color information is very useful to identify quasars and distinguish them from contaminants.
This photometry also contains useful redshift information; recent approaches to estimate redshifts from photometry with neural networks achieve a mean $|\Dz|\sim 0.22$ \citep{yang_quasar_2017, jin_efficient_2019, kunsagi-mate_photometric_2022}.
In our case of redshift estimates from narrow-range BP/RP spectra, we expect IR photometry to add information that can break line identification degeneracies in order to improve estimates.
We incorporate the $W1$ and $W2$ bands into both our quasar selection (\S\ref{sec:decontam}) and redshift estimation (\S\ref{sec:redshifts}) procedures.

\subsection{\SDSS DR16 quasar sample}
\label{sec:data_sdss_quasars}

The Sloan Digital Sky Survey released the largest spectroscopic quasar catalog in DR16\footnote{The \SDSS DR16Q quasar catalog is publicly available at \url{https://www.sdss.org/dr16/algorithms/qso_catalog}.} \citep{lyke_sloan_2020}.
It combines new sources from the extended Baryon Oscillation Spectroscopic Survey (eBOSS), part of \textsl{\SDSS-IV}, with previously observed sources from earlier \SDSS campaigns.
The catalog contains 750,414 quasars, with an estimated 99.8\% completeness (compared to the \SDSS-III/SEQUELS sample of \citealt{myers_sdss-iv_2015}, which has higher signal-to-noise spectra) and $98.7$--$99.7$\% purity.
We remove sources with redshift warnings, \texttt{ZWARNING}!=0, as well as a handful of sources with unreasonably low or negative redshift estimates ($z<0.01$). 
This results in \val{N_sqall} sources, which is the sample shown in Figure~\ref{fig:frac_matrix}.
We crossmatch these with the \Gaia catalog, as well as \unWISE (\S\ref{sec:data_wise}), using a maximum separation of 1\arcsec on the sky.
We remove sources with fewer than five observations in $BP$ (\texttt{phot\_bp\_n\_obs}) or $RP$ (\texttt{phot\_rp\_n\_obs}), following \cite{bailer-jones_dsc_2021}, as well as sources that are duplicated in the \SDSS star or galaxy samples (\S\ref{sec:data_contaminants}). 
This results in \val{N_squasars_unwise} sources with both \Gaia and \unWISE observations that pass these criteria.

We use these to calibrate the cuts to decontaminate our sample (\S\ref{sec:decontam}); for this purpose, we only keep sources that are also in the \cat superset (sources that are in the \Gaia quasar candidates table, have all necessary \Gaia and \unWISE photometry, \Gaia-estimated QSOC redshifts, and $G < \Gmax$).
This sample contains \val{N_squasars_sup} quasars.
We also use this sample (after applying the cuts described in \S\ref{sec:decontam}) to train our redshift estimation model (\S\ref{sec:redshifts}).
While this spectroscopic sample has quite high completeness and accurate redshift information, we note that it is still imperfect, contains selection effects, and represents only a particular definition of a quasar; these issues will propagate to our catalog.

\subsection{Contaminant samples: galaxies and stars}
\label{sec:data_contaminants}

To guide the decontamination of our catalog (\S\ref{sec:decontam}), we compile known contaminant samples, namely galaxies and stars.
For the galaxy sample, we use \SDSS spectroscopic galaxies from DR18\footnote{\SDSS DR18 data can be accessed at \url{https://skyserver.sdss.org/CasJobs/jobdetails}.}.
Following \cite{bailer-jones_dsc_2021}, we include all galaxies with class label \texttt{GALAXY} in the \texttt{SpecObj} table, exclude galaxies with subclass labels \texttt{AGN} or \texttt{AGN BROADLINE}, and exclude sources with redshift warnings, \texttt{zWarning=0}.
We crossmatch these with \Gaia DR3 and \unWISE with a 1\arcsec radius, and remove sources with fewer than five observations in $BP$ or $RP$, as for the \SDSS quasars.
We also remove apparent stellar contaminants from the galaxies sample with the cut in $G$--$RP$ and $BP$--$G$ from equation (1) of \cite{bailer-jones_quasar_2019}, and additionally remove sources duplicated in the \SDSS quasar or star samples.
This leaves \val{N_sgals_unwise} crossmatched \SDSS galaxies in our sample; \val{N_sgals_sup} of these are in the \cat superset.

For the star sample, we also use \SDSS DR18 sources, selecting objects with class label \texttt{STAR} in the \texttt{SpecObj} table.
As for the quasars and galaxies, we crossmatch these with \Gaia DR3 with a 1\arcsec radius and remove sources with fewer than 5 observations in BP or RP, and remove sources duplicated in the other samples.
This results in a stellar sample with \val{N_stars_unwise} crossmatched \SDSS-\Gaia stars, with \val{N_sstars_sup} of these in the superset.

For the decontamination procedure, we also compile a sample of sources in or near the LMC or SMC, as most of these will be stellar contaminants but have different properties than the \SDSS star sample.
To do this, we select all sources in the \Gaia quasar candidates table that are within 3\textdegree\ of the center of the LMC or 1.5\textdegree\ from the center of the SMC.
While this may include stars not actually in the LMC or SMC, we have chosen these fairly narrow radii in order to capture mostly LMC and SMC stars and few potential quasars.
Additionally requiring that these have \unWISE photometry, this gives \val{N_mcs_unwise} LMC- and SMC- adjacent stars; \val{N_mcs_sup} are in the superset.

\section{Catalog construction}
\label{sec:construction}

\subsection{Decontamination with proper motions and \unWISE colors}
\label{sec:decontam}

\begin{figure}
    \centering
    \includegraphics[width=0.9\columnwidth]{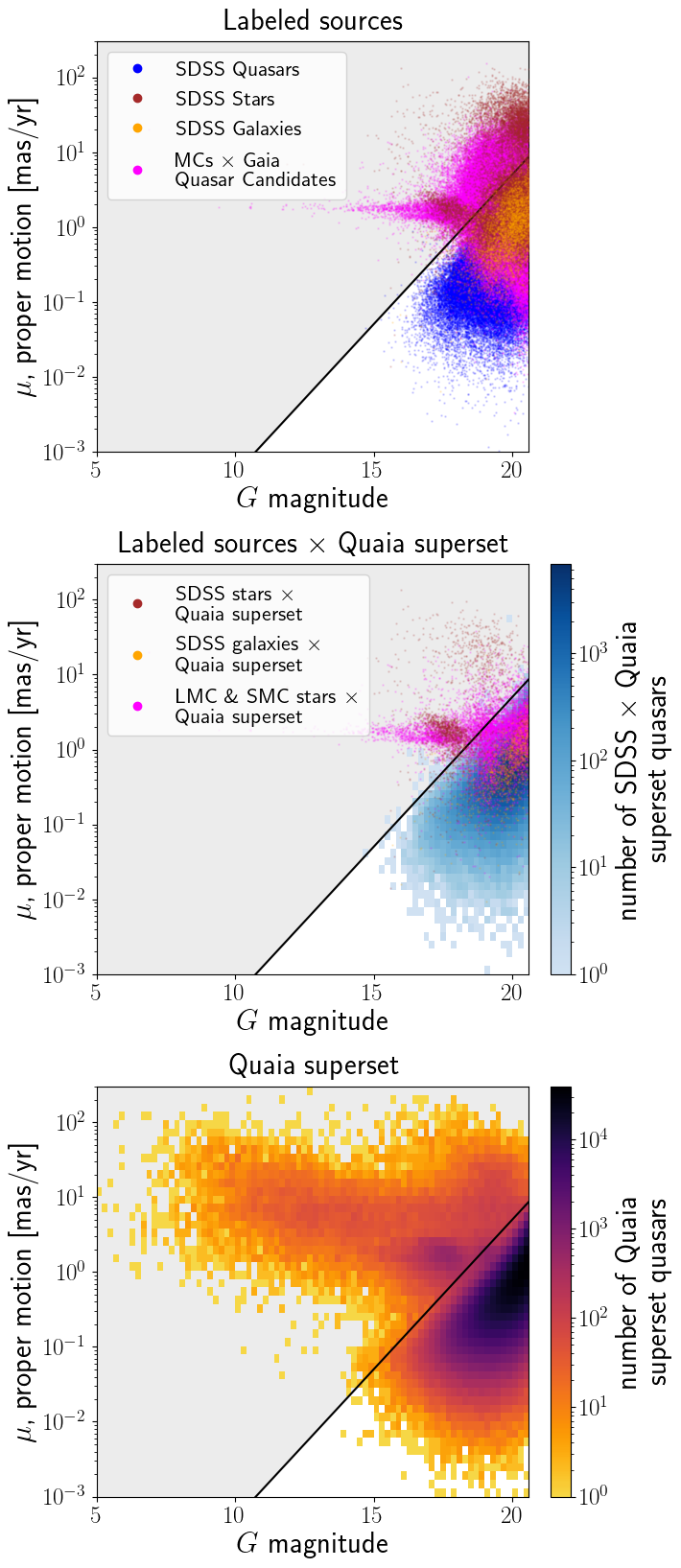}

    \caption{Proper motion $\mu$ vs. $G$ magnitude  for two different sets of sources. The black line shows the cut we make; the shaded gray region is excluded from the catalog. \emph{Top:} the sources for which we have labels (\SDSS data as well as sources near the LMC and SMC in the \emph{Gaia} quasar candidates sample) that are also in the \cat superset (\Gaia DR3 quasar candidates that have all necessary photometry, \Gaia redshift estimates, and $G<\Gmax$). \emph{Middle:} sources in the top row that are also in the \cat superset. \emph{Bottom:} the superset of quasar candidates from which \cat is constructed. The proper motion cut includes nearly all \SDSS quasars in the superset while excluding a large number of stars.} 
    \label{fig:G_pm}
\end{figure}

The full \Gaia quasar candidate sample is known to contain a significant fraction of contaminants (stars and other non-quasars, such as galaxies).
The stellar contaminants might include sources such as brown dwarfs, which have similar colors as high-redshift quasars, and potentially blue horizontal branch stars, blue stragglers, and white dwarfs, which are UV bright like lower-redshift quasars.
To remove stellar contaminants, we make an initial cut on proper motion $\mu$, as quasars should have negligible proper motions due to their large distances.
The value of $\mu$ has a dependence on $G$, so we make a cut in this space.
To guide this cut, we use labeled sources: \SDSS quasars, \SDSS galaxies, \SDSS stars, and \Gaia LMC- and SMC-adjacent stars, as described in \S\ref{sec:data_sdss_quasars} and \S\ref{sec:data_contaminants}.
The $G$--$\mu$ distributions of these sources are shown in the top panel of Figure~\ref{fig:G_pm}.
In the middle panel, we show the intersection of these labeled sources with our \cat superset, which consists of sources in the \Gaia quasar candidates table that have \Gaia redshift estimates, complete \Gaia and \unWISE photometry, and are below $G<\Gmax$.
We see that the \SDSS quasars tend to have much smaller proper motions than the other types of sources, with a very linear edge to the $G$ dependence at the high proper motion side of the distribution.
Based on this, we choose the cut
\begin{equation}
    \mu < 10^{0.4\,(G-18.25)} \: \text{mas/yr} ~.
\end{equation}
At $G=18.25$, this corresponds to $\mu \lesssim 2.5 \: \text{mas}\:\text{yr}^{-1}$, and allows for less severe cuts at deeper magnitudes given the typically less precise astrometry.
This is related to the proper motion \emph{uncertainty} as a function of $G$, which has been quantified by \Gaia \citep{gaia_collaboration_gaia_2021}.
We show this cut overlaid on the \cat superset in the lower panel of Figure~\ref{fig:G_pm}; based on the labeled data, we can clearly pick out the populations.
The proper motion cut excludes \val{N_removed_pmcut} sources, $\val{p_removed_pmcut}\%$ of the superset.

\begin{figure*}[p]
    \centering
    \includegraphics[width=0.75\textwidth]{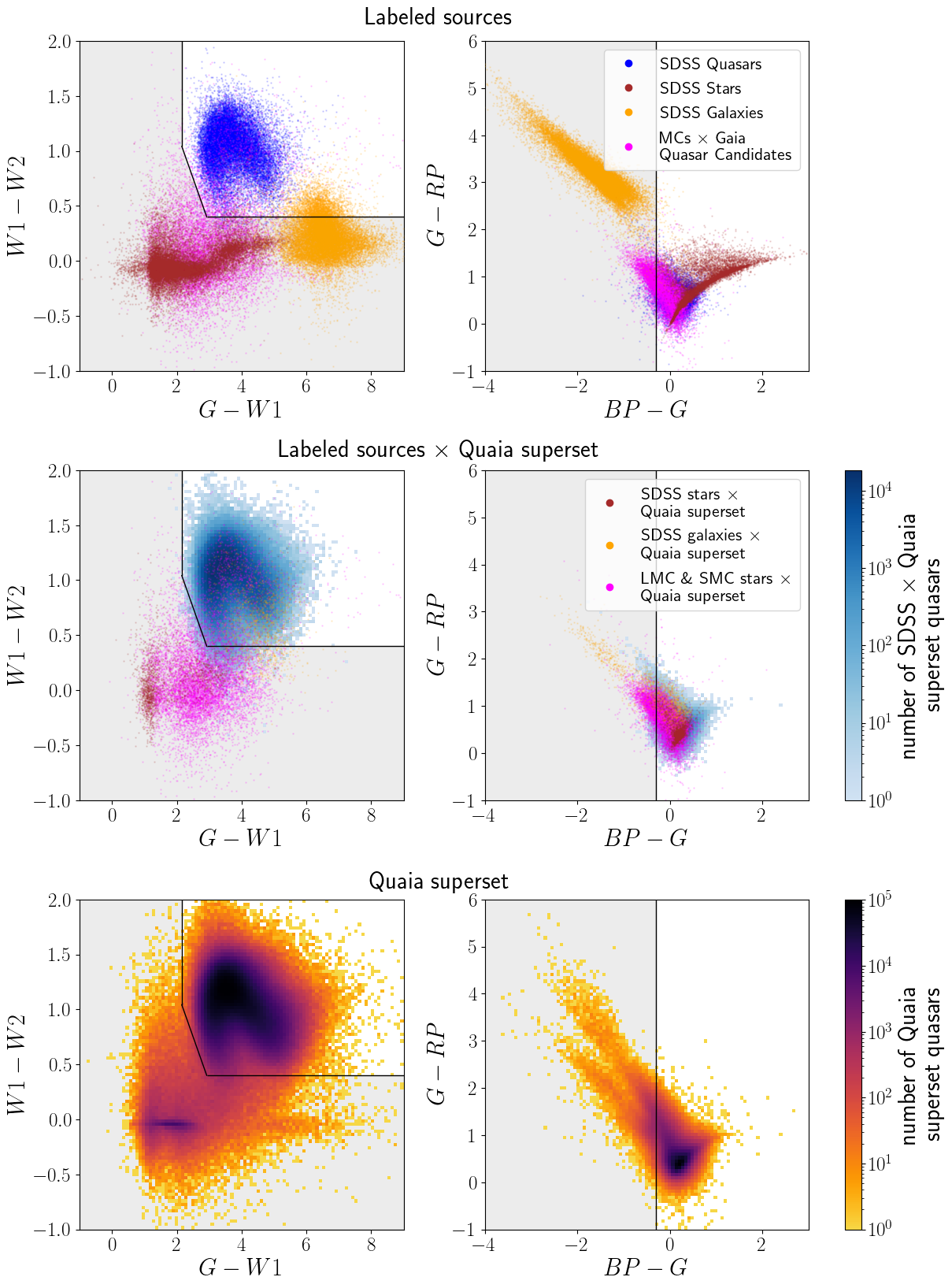}

    \caption{Color--color plots of three different sets of sources. The left column shows $W1$--$W2$ vs. $G$--$W1$ color, and the right column shows $G$--$RP$ vs. $BP$--$G$ color. The black lines show the cuts we make; the shaded gray region is excluded from the catalog. The rows have the same samples as in Figure~\ref{fig:G_pm}, except that in the top row, only 20,000 of each type of \SDSS source is shown for clarity. In both color-color projections, the labeled sources are mostly localized in particular regions of parameter space, and we can see these populations somewhat clearly in the \cat superset.} 
    \label{fig:color_color}
\end{figure*}

Next, we determine the color cuts based on \Gaia and \unWISE photometry.
Generally, stars and galaxies are dim in redder, IR wavelengths compared to AGN.
For instance, the eBOSS \new{quasar target selection \citep{myers_sdss-iv_2015}} involved linear cuts in the optical-IR, involving the \SDSS $g$, $r$, and $i$ bands and the WISE $W1$ and $W2$ bands. 

In Figure~\ref{fig:color_color}, we show color-color distributions for the same samples as in Figure~\ref{fig:G_pm}.
The left panel shows $W1$--$W2$ vs. $G$--$W1$ color, and the right column shows $G$--$RP$ vs. $BP$--$G$ color.
The top row, with the full labeled samples, shows that different types of sources tend to be localized to different areas of this parameter space (we show only a subset of each type for clarity).
In particular, the colors involving \unWISE (left panel) separate out the source types relatively clearly, demonstrating the importance of the \unWISE crossmatch: \SDSS quasars have very red $W1$--$W2$, and intermediate $G$--$W1$ color, while galaxies have bluer $W1$--$W2$ and redder $G$--$W1$ compared to quasars, and stars (both \SDSS stars and stars near the LMC and SMC) are bluer in both colors.
In \Gaia color-color space, galaxies tend to have bluer $BP$--$G$ and redder $G$--$RP$ colors than the other types of sources.
In the middle row of Figure~\ref{fig:color_color}, showing the intersection of the labeled sources with the \cat superset, we see that the superset restrictions have eliminated many of the sources, especially \SDSS galaxies and stars, though a significant number remain.
(We note that it is possible that some of these \SDSS galaxies do host AGN though they were not classified as such by \SDSS.)
The \cat superset is shown in the bottom panel; we can see clear populations of quasars, stars, and galaxies lining up with the labeled sources.
Importantly, we can see the effect of the stricter \SDSS color selection in the red (high $G$--$W1$) region of parameter space into which the \Gaia quasar candidates extend, but are not represented in the \SDSS sample in the above panels.

We choose to apply linear cuts in these colors to decontaminate the sample.
While other works (e.g. \citealt{hughes_quasar_2022}) train classifiers to determine which objects are true quasars \new{using \SDSS-classified quasars as labels}, we opt for simpler cuts for ease of reproducibility and to \new{mitigate the propagation of} \SDSS selection effects, \new{which may include color- and magnitude-dependent effects}.
We choose four cuts based on the distribution of sources in color-color space. 
The first is in $W1$--$W2$, which has been shown to be useful for distinguishing quasars; for instance, \cite{nikutta_meaning_2014} demonstrated that a small crossmatched \SDSS quasar sample has very red $W1$--$W2=1.2 \pm 0.16$, while other types of objects---namely star-forming and AGN galaxies, luminous red galaxies and stars---have bluer $W1$--$W2$.
Stars tend to have the bluest $W1$--$W2$, with a mean of $W1$--$W2=-0.04 \pm 0.03$, so a cut in $W1$--$W2$ is a reliable way to filter out stellar contaminants.
We add a cut in $G$--$W1$ to filter out the bulk of the stars (including the LMC and SMC), and another in $BP-G$ to cut out the galaxy contaminants.
Finally, we find that these single color cuts were not sufficient to remove all of the LMC and SMC, so we add an additional diagonal cut in $W1$--$W2$ and $G$--$W1$, choosing a reasonable slope.

We optimize the values (intercepts) of these four cuts with a grid search, trying values spaced out by 0.1 mag.
We note that while we show the full samples in Figure~\ref{fig:color_color}, in practice we make the proper motion cut before optimizing the color cuts.
We choose the color cuts that maximize our objective function $\mathcal{L}$,
\begin{equation}
    \mathcal{L} = N_\text{q} - \lambda_\text{s} \, N_\text{s} - \lambda_\text{g} \, N_\text{g} - \lambda_\text{m} \, N_\text{m} ~,
\end{equation}
where $N_\text{q}$ is the number of true quasars that make it into the catalog, $N_\text{s}$ \SDSS stars, $N_\text{g}$ \SDSS galaxies, and $N_\text{m}$ LMC and SMC stars, and the $\lambda$ parameters balance the relative ratios of each.
We choose $\lambda_\text{s}=3$, $\lambda_\text{m}=5$, and $\lambda_\text{g}=1$.

The optimal cuts for the objects to keep in the catalog are
\begin{equation}
\begin{split}
    (G-W1) &> 2.15 \\ (W1-W2) &> 0.4 \\ (BP-G) &> -0.3 \\ (G-W1) + 1.2\,(W1-W2) &> 3.4
\end{split}
\end{equation}
These are shown as the black lines in all panels of Figure~\ref{fig:color_color}, with the gray shading indicating exclusion regions.
These cuts, as well as the proper motion cuts described above, exclude ${\sim}\val{p_cut_gsup_gclean}\%$ of the superset, resulting in \val{N_gclean} quasars in our \textit{decontaminated} sample.
We apply an additional magnitude cut of $G<\Ghi$ to reduce edge effects in our redshift estimation; this constitutes our deep sample, with \val{N_gcathi} sources.
We refer to this as \cat in the rest of this work.
However, the catalog becomes less clean and reliable as we push to deeper magnitudes---due to less precise measurements and stronger systematics, notably the \Gaia scanning pattern---so we produce a version of the catalog with $G<\Glo$ to ensure a cleaner sample.
This brighter catalog has \val{N_gcatlo} sources, and we report most of our results on this sample throughout the rest of this work.

\subsection{Spectrophotometric redshifts with \unWISE and \SDSS}
\label{sec:redshifts}

\begin{figure*}
    \centering
    \subfloat[\label{fig:zgaia_zsdss}]{\includegraphics[width=0.45\textwidth]{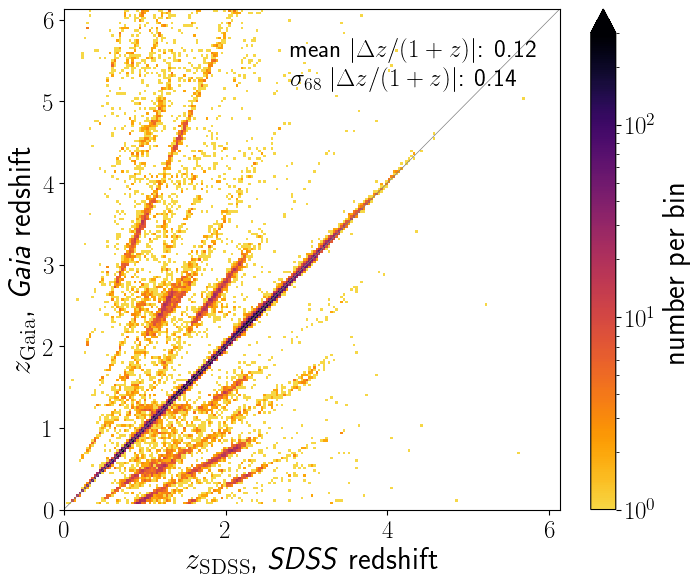}}
    \hspace{5ex}
    \subfloat[\label{fig:zspz_zsdss}] {\includegraphics[width=0.45\textwidth]{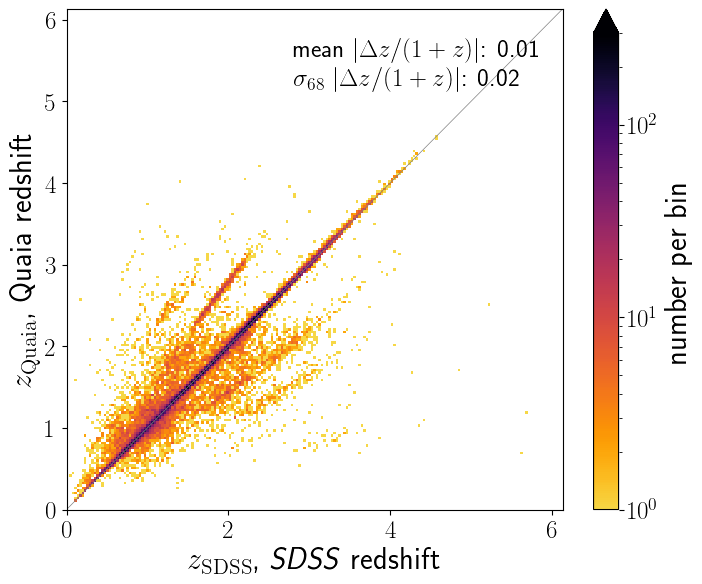}}
    \caption{(a) \Gaia redshift estimate $\zgaia$ vs. \SDSS (``true'') redshift $\zsdss$ for a test set of sources in our quasar catalog \cat with $G<\Ghi$. (b) Our estimated spectrophotometric (SPZ) redshifts $\zquaia$, \new{which are based on a \knn model,} vs. $\zsdss$ for the same sample. \new{The bias (mean redshift error) and scatter ($\sigma_{68}$, the symmetrized inner 68\% region of the redshift errors) of the redshift estimates compared to $\zsdss$ are shown in the panels.} The $\zquaia$ redshifts significantly decrease both the bias and scatter, as well as catastrophic outliers and unreasonably high redshift estimates. The one-to-one line (perfect accuracy) is shown in gray; note that the color bar is on a log scale, and that a majority of the sources in both cases lie along this line.}
    \label{fig:zsdss_comp}
\end{figure*}

We use \unWISE and \SDSS data to improve the redshift estimation of the sources.
Figure~\ref{fig:zgaia_zsdss} shows the redshifts estimated by the \Gaia QSOC pipeline $\zgaia$ compared to the \SDSS redshifts $\zsdss$ for a test sample of sources from \cat with $G<\Ghi$; note that the 2D histogram is plotted in log-space to show the outliers more clearly.
We find that of the \Gaia redshifts $\zgaia$, $\val{p_acc_zgaia_dzhi_Glo}\%$ ($\val{p_acc_zgaia_dzmid_Glo}\%$) agree to $|\dz|<\val{dzhi}$ ($\val{dzmid}$).
A significant fraction of $\zgaia$ are highly precise: $\val{p_acc_zgaia_dzlo_Glo}\%$ agree with \SDSS to $|\dz|<\val{dzlo}$.
We also clearly see bands of incorrect estimation due to line aliasing issues.
Additionally, in the crossmatched sample, nearly all of the very high $\zgaia$ estimates ($z>4.5$) are shown to be incorrect in comparison to \SDSS.
We note that the redshift estimation is much more accurate for sources that have no redshift warning flags set (\texttt{flags\_qsoc=0}), as discussed in \S\ref{sec:data_gaia}, but this is only true for $\val{p_gcathi_flagsqsoc0}\%$ of the sources in \cat ($G<20.5$), and even including sources with \texttt{flags\_qsoc}=16 this leaves only $\val{p_gcathi_flagsqsoc0or16}\%$ of sources.

We train a \knn model on \cat sources to estimate improved redshifts.
(We also tested other models including XGBoost and a multilayer perceptron, and found that the \knn outperformed both by a small margin.)
We include all sources in our decontaminated catalog (\S\ref{sec:decontam}) which goes out to $G<\Gmax$, in order to have a buffer beyond our desired $G<\Ghi$ sample to reduce edge effects from the training set.
(We find that including the rest of the photometry does not make a difference in the results.)
The reddening is determined with the Corrected Schlegel, Finkbeiner, \& Davis (SFD) dust map introduced by \new{\cite{chiang_corrected_2023}, which corrects the standard \cite{schlegel_maps_1998} dust map by subtracting off the contribution from the cosmic infrared background (CIB).
(We also include the appropriate correction factor given by \cite{schlafly_measuring_2011}.)}\footnote{The dust map was accessed with the Python package \url{https://dustmaps.readthedocs.io}}.
The labels are the \SDSS redshifts, $\zsdss$.

We use as our labeled data sources from the crossmatched \SDSS DR16Q sample (\S\ref{sec:data_sdss_quasars}) that are also in our decontaminated catalog \cat, so that we train on sources drawn from the same distribution to which we will apply the model; this is \val{N_sqclean} sources.
We apply a 70\%/15\%/15\% train/validation/test split.
We build a $k$-d tree on the training set features using the \texttt{KDTree} implementation of \texttt{sklearn}.
At the prediction stage, we access the $K$ nearest neighbors of each input feature vector, first excluding neighbors with zero distance in feature space (i.e. neighbors that are in the training set).
We assign the predicted label to be the median $\zsdss$ of the $K$ nearest neighbors, and the uncertainty to be the symmetrized inner 68\% error of those neighbors.
We use the validation set to tune $K$, and choose the value that maximizes the fraction of predicted redshifts with $|\dz|<\val{dzmid}$, which is $K=27$; we note that this value only varies at the $\sim$1\% level for values $15 < K < 50$, and is similar for other choices of $|\dz|$. 
Finally, we apply the model to the full \cat and output \knn redshift estimates, $\zknn$, for each source.

\begin{figure}
    \centering
    \includegraphics[width=\columnwidth]{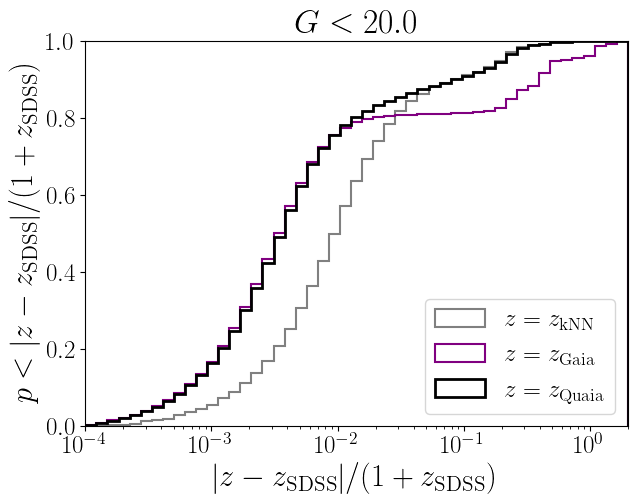}
    \caption{The cumulative distribution of redshift errors for \cat test set sources with $G<\Glo$, considering \SDSS spectroscopic redshifts $\zsdss$ as the ground truth, for estimates directly from our \knn model (gray), the original $\zgaia$ redshifts (purple), and our final $\zquaia$ estimates (black) based on a combination of the other two. Our SPZ redshifts have far fewer outliers and similar precision compared to the \Gaia estimates.}
    \label{fig:z_error_cumulative}
\end{figure}

The results are shown in Figure~\ref{fig:z_error_cumulative}, which shows the cumulative distribution of errors $|\dz|$ for $\zknn$ compared to that of $\zgaia$ (with $\zsdss$ as the truth) for the test set with $G<\Glo$.
(The shapes are similar for $G<\Ghi$, just shifted to somewhat lower accuracy.)
We find that the $\zknn$ estimates have far fewer outliers than $\zgaia$.
However, the $\zgaia$ estimates tend to be more precise, as they use the full spectral information, while the \knn is essentially smoothing over the likeliest neighboring sources in feature space. 
We thus choose to combine the properties of both of these redshift estimates to obtain our final \emph{spectrophotometric} (SPZ) redshifts $\zquaia$ in the following way.
For sources for which $\zquaia$ and $\zgaia$ agree to $|\dz|<0.05$, we assign $\zquaia = \zgaia$ to preserve the precision of the \Gaia estimate.
For sources for which $\zquaia$ and $\zgaia$ differ by $|\dz|>0.1$, we assign $\zquaia = \zknn$ to preserve accuracy.
In between these thresholds, we apply a smooth, linear transition to avoid hard features in our estimates.
These $\zquaia$ estimates are also shown in Figure~\ref{fig:z_error_cumulative} compared to the ``true'' (spectroscopic, taken as truth for our purposes) \SDSS redshifts, and we can see that these achieve nearly as high precision as $\zgaia$ while maintaining the high accuracy of $\zknn$.

Our $\zquaia$ results for the test set are shown in Figure~\ref{fig:zspz_zsdss} compared to $\zsdss$, shown here for the full catalog depth $G<\Ghi$. 
We find that $\val{p_acc_zspz_dzhi_Ghi}\%$ ($\val{p_acc_zspz_dzmid_Ghi}\%$) of our SPZ redshifts agree to $|\dz|<\val{dzhi}$ $(\val{dzmid})$, and $\val{p_acc_zgaia_dzlo_Ghi}\%$ highly agree to $|\dz|<\val{dzlo}$.
We also give the \new{bias (mean redshift error) and scatter ($\sigma_{68}$, the symmetrized inner 68\% region of the redshift errors)} of $|\dz|$ in the figure; our SPZ redshifts significantly decrease the bias and scatter.
The SPZ estimation corrected all of the very high-$z$ \Gaia estimates, and some of the intermediate-outlying aliasing effects.
We still have some catastrophic outliers due to line aliasing, but with our SPZ redshifts, we find a reduction in the number of $|\dz|>\val{dzhi}$ (\val{dzmid}) outliers by \val{factor_reduction_outliers_dzhi_Ghi} (\val{factor_reduction_outliers_dzmid_Ghi}) compared to the \Gaia redshift estimates.

\begin{figure}
    \centering
    \includegraphics[width=\columnwidth]{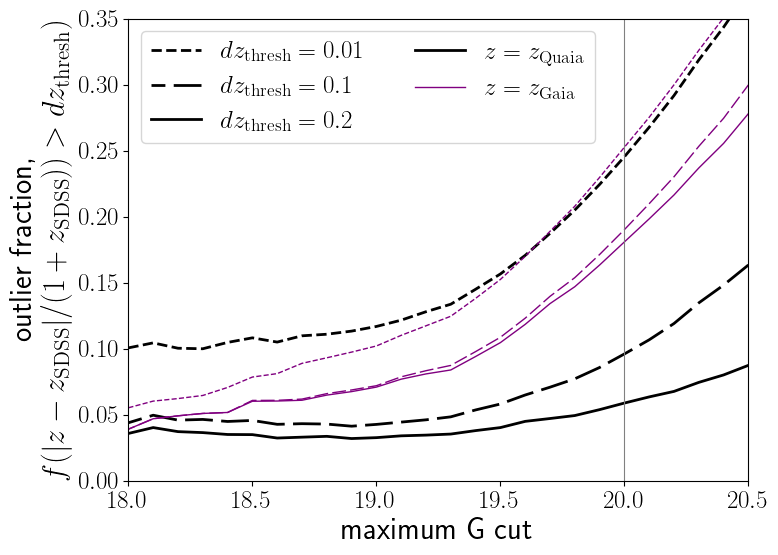}
    \caption{The fraction of outlying redshifts with $|\dz| > (\val{dzlo}, \val{dzmid}, \val{dzhi})$, as a function of $G$ magnitude, for our redshift estimation test set. The SPZ redshifts are shown in black, and the \Gaia redshifts in purple. The fraction of outliers increases steeply with increasing $G$ for $G>19.5$ for both $\zquaia$ and $\zgaia$, though the fraction of catastrophic outliers for $\zquaia$ is significantly lower (and the dependence less steep) compared to $\zgaia$.}
    \label{fig:z_G_dep}
\end{figure}

We investigate the dependence of the redshift error on the $G$-band magnitude in Figure~\ref{fig:z_G_dep}.
The fraction of redshifts with an error above various thresholds is shown as a function of samples with the given cut on $G$.
The errors are lowest at a bright magnitude cut of $G < \sim$$\val{Gbright}$; in this sample, sources with SPZ redshift estimates inaccurate to $|\dz|>\val{dzhi}$ (\val{dzmid}) comprise only $\val{p_outliers_zspz_dzhi_Gbright}\%$ ($\val{p_outliers_zspz_dzmid_Gbright}\%$) of the sample, and to the more stringent requirement of $|\dz|>\val{dzlo}$, $\val{p_outliers_zspz_dzlo_Gbright}\%$.
This outlier fraction increases steadily as fainter sources are included.
For $G<\Glo$, $\val{p_outliers_zspz_dzhi_Glo}\%$ ($\val{p_outliers_zspz_dzmid_Glo}\%$) are inaccurate to $|\dz|>\val{dzhi}$ $(\val{dzmid})$, and $\val{p_outliers_zspz_dzlo_Glo}\%$ for $|\dz|>\val{dzlo}$.
Compared to the \Gaia redshift estimates, the SPZ estimates $\zquaia$ reduce the number of $|\dz|>\val{dzhi}$ (\val{dzmid}) outliers by \val{factor_reduction_outliers_dzhi_Glo} (\val{factor_reduction_outliers_dzmid_Glo}).
The choice of $G$ cut to use in a given analysis will depend on the nature of the analysis and its sensitivity to outliers. 

We note that our finding that the \unWISE IR information significantly improves redshift estimates, compared to only the optical information used in the \Gaia QSOC estimates, is consistent with other photometric redshift work.
\new{For instance, \cite{dipompeo_quasar_2015} showed that including WISE mid-IR photometry in the redshift estimation of \SDSS-imaged quasars results in a significant improvement on the estimates, even more so than including both GALEX near- and far-UV data and UKIDSS near-IR data.
More recently,} \cite{yang_southern_2023} compiled a photometric quasar catalog from the Dark Energy Survey (DES) DR2, combining DES optical photometry with near-IR photometry as well as \unWISE mid-IR photometry; they obtained photo-$z$s with 92\% having $|\dz|<0.1$ when all IR bands are used compared to 72\% with only optical data.
Additional photometric information at other wavelengths could further improve our estimates \new{(as well as catalog decontamination)}, but is not currently available for enough sources in our \cat catalog to be worthwhile.
\new{For instance, for the UV all-sky survey GALEX \citep{martin_galaxy_2005}, crossmatches to Quaia sources are only available for $\val{p_nuv_galex}\%$ of the \cat objects for near-UV observations, and when including far-UV only \val{p_nuv_fuv_galex}\%; this significant discrepancy is largely due to the faint end of Quaia, where GALEX observations do not reach deep enough.}
\new{The Pan-STARRS1 survey \citep{chambers_pan-starrs1_2019} covers only three quarters of the sky, with crossmatches to $\val{p_panstarrs}\%$ of Quaia sources.
We tested adding Pan-STARRS1 data to the redshift estimation feature set and found only a small improvement, and thus chose to prioritize keeping the full sky span of \cat, though we make note that incorporating Pan-STARRS1 may be useful for certain applications.}

\subsection{Selection Function Modeling}
\label{sec:selfunc_methods}

\begin{figure*}
    \centering

    \subfloat[\label{fig:systematics_dust}]{\includegraphics[width=0.45\textwidth]{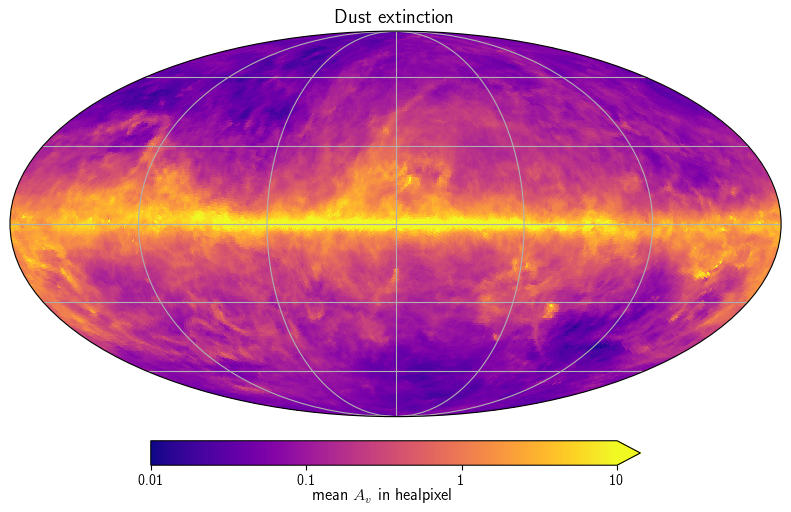}} \\

    \subfloat[\label{fig:systematics_stars}]{\includegraphics[width=0.45\textwidth]{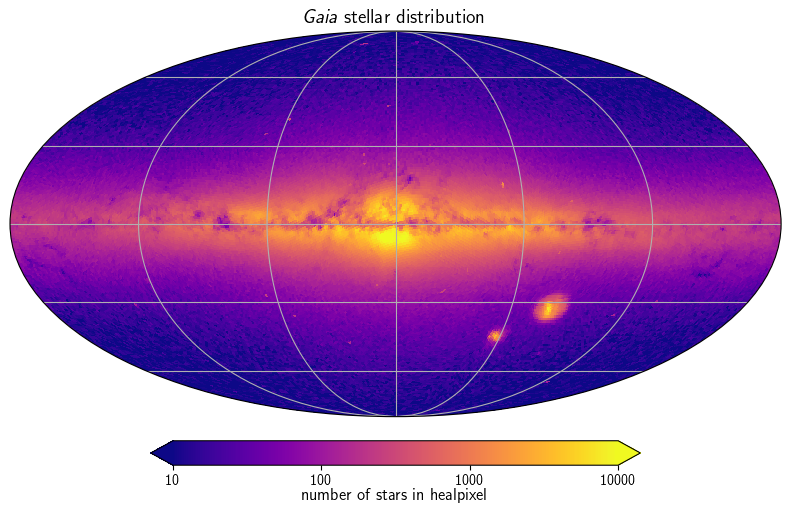}}
    \hspace{5ex}
    \subfloat[\label{fig:systematics_unwise}]{\includegraphics[width=0.45\textwidth]{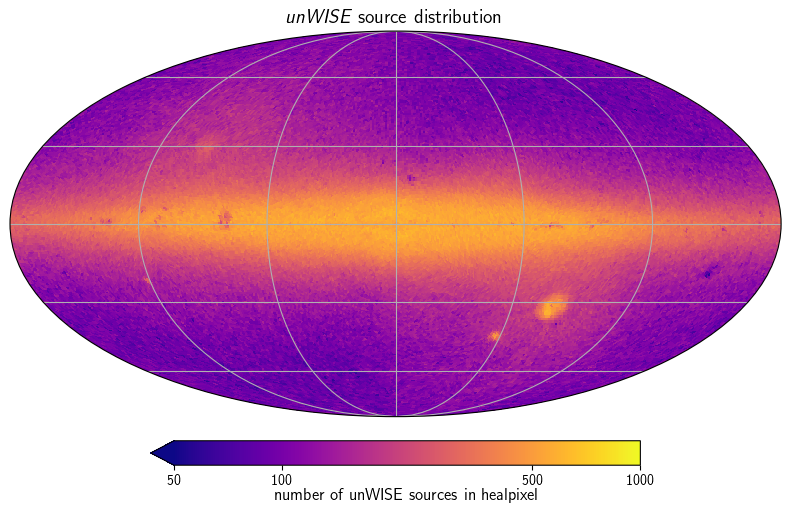}}

    \subfloat[\label{fig:systematics_m10}]{\includegraphics[width=0.45\textwidth]{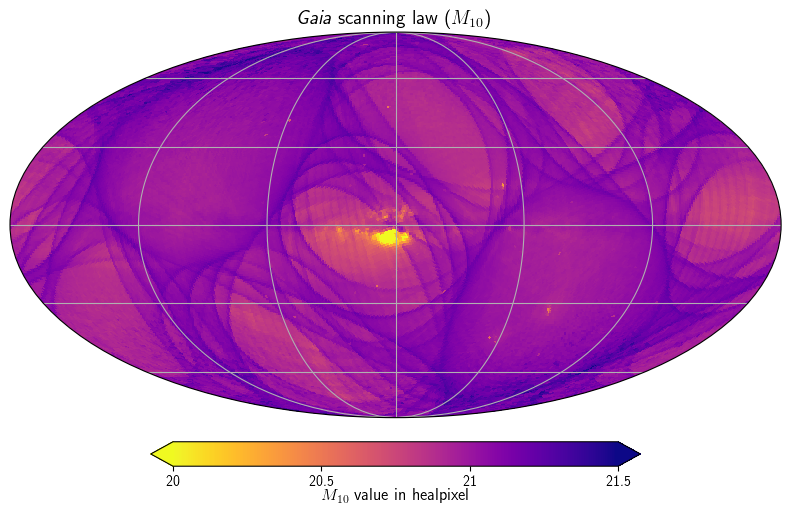}}
    \hspace{5ex}
    \subfloat[\label{fig:systematics_unwisescan}]{\includegraphics[width=0.45\textwidth]{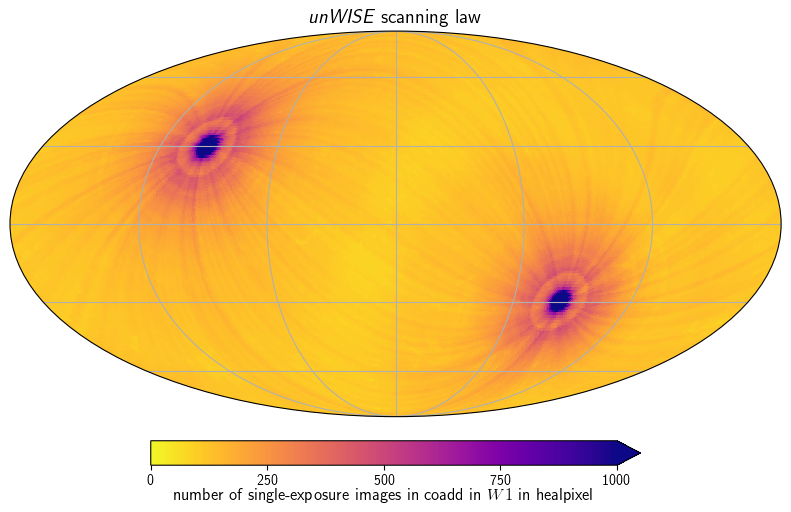}}
    

    \caption{The systematics maps used in the selection function model: (a) dust extinction \new{from \cite{chiang_corrected_2023}}; (b) the stellar distribution based on $\sim$10.6 million randomly selected \Gaia sources with $18.5 < G < 20$; \new{(c) the \unWISE source distribution based on $\sim$10.6 million randomly selected \unWISE sources;} (d) the quantity $M_{10}$, the median magnitude of sources with $\leq10$ \Gaia transits, which encodes the \Gaia scanning law and source crowding; \new{and (e) the \unWISE scan pattern given by the mean number of single-exposure images of the sky region in the coadd}. Note that the color bar on the $M_{10}$ \new{and \unWISE scanning law maps} are reversed, as high \new{values} indicate a cleaner region, the inverse of the other maps.
    \new{We also include separate templates for only sources in the LMC and SMC regions for both the stellar and \unWISE source densities, with the background subtracted. All templates are discussed in more detail in the text.}
    }
    \label{fig:systematics}
\end{figure*}

Observational and astrophysical effects impact which sources we observe and their properties; this is known as the selection function. 
As \Gaia is a space-based mission, it avoids many of the observational issues of ground-based surveys, such as seeing and airmass.
However, there are still significant selection effects: for our model, we consider dust, \new{the source density of the parent surveys}, and \new{the scan patterns of the parent surveys}.

We fit a selection function model to a particular version of the catalog, namely, a particular maximum $G$.
For the fiducial selection function we work only in terms of sky position.
We make a healpix map of the catalog with NSIDE = 64 and count the number of observed catalog sources in each healpix pixel.
\new{We choose this NSIDE, which results in 49,152 pixels each with an area of $\sim$0.84 deg$^2$, to balance constructing a map with reasonably high resolution with ensuring a sufficient number of sources in the pixels for stable fits, as well as fitting within memory limitations for the Gaussian process fit.}
In the case of no selection effects (and under the assumption of isotropy), we would expect each pixel to contain roughly the same number of sources.
Our goal is to model the dependence between the number of sources per pixel and the various systematics.

The systematics maps (templates) we use are shown in Figure~\ref{fig:systematics}. 
We use the dust map of \new{\cite{chiang_corrected_2023}}, and convert it to a healpix map of NSIDE = 64.
To do this, we evaluate the reddening $E(B-V)$ at the centers of pixels of a high-resolution NSIDE = 2048 healpixelization of the sphere\new{, and apply the 0.86 correction factor proposed by \cite{schlafly_measuring_2011}}.
We convert these to extinction values by multiplying by $R_V=3.1$, and then take the mean of all of these values within each healpixel target NSIDE = 64 map.
This produces a smoothed dust extinction map on the desired scale.
The result is shown in Figure~\ref{fig:systematics_dust}; the extinction is highest around the Galactic plane, with structure extending outward.

For the stellar distribution, we randomly select $\sim$10.6 million \Gaia sources with $18.5<G<20$, the magnitude range of most of our quasar sample.
The vast majority of these will be true stars.
(While this sample will contain some other types of objects, including possibly some quasars and other extragalactic sources, these will be orders of magnitude less numerous than stars.)
We count the number of stars per NSIDE = 64 healpixel; this is shown in Figure~\ref{fig:systematics_stars}.
\new{We also include a template of the \unWISE source distribution, for which we randomly selected $\sim$10.6 million \unWISE sources (1\% of the catalog) that have flux in both $W1$ and $W2$, and have primary status (\texttt{Prim}=1).
We count the number of these sources per NSIDE = 64 healpixel}, as shown in Figure~\ref{fig:systematics_unwise}.

\new{In initial fits we found that the regions of the LMC and SMC are particularly poorly modeled, and that the fit is improved by including separate templates of just the LMC and SMC source density for both the \Gaia and \unWISE sources; this gives the model the freedom to assign different coefficients to these regions than to the overall survey source density.
(The need for different coefficients could be for a number of reasons, such as a difference in stellar density, contamination, or magnitude distribution; we leave a deeper investigation of this to future work and just use this empirical finding to improve our model.)
For the LMC/SMC templates,} we cut out a wide region around the LMC and SMC (9\textdegree\ in radius around the LMC and 5\textdegree\ around the SMC), and subtract the background, which we approximate using the region at the same latitude but opposite longitude (mirrored across the $l=0$\textdegree\ line) of the \new{given source} distribution map.
We don't show these maps here as they are visually similar to the stellar and \unWISE source density maps in the LMC and SMC regions (though with the background subtracted).

For the \Gaia completeness, we use the quantity $M_{10}$ introduced by \cite{cantat-gaudin_empirical_2023}\footnote{This map can be accessed with the \texttt{gaiaunlimited} package, \url{https://gaiaunlimited.readthedocs.io}}.
$M_{10}$ is the median magnitude in a given sky patch of the \Gaia sources with $\leq10$ transits across the \Gaia field of view; it incorporates the effects of both the scanning law and source crowding.
The actual completeness map derived by \cite{cantat-gaudin_empirical_2023} depends on both $M_{10}$ and $G$-band magnitude; this completeness is very close to 1 for nearly all of the sky for $G=\Glo$, with some non-negligible incompleteness for $G=\Ghi$.
However, this completeness model is based on the full \Gaia source catalog, while we expect the selection function of our quasar sample to be different.
We thus use the $M_{10}$ map directly in our fit to capture the effects of the \Gaia scanning law and source crowding specific to \cat.
We downsample the map to NSIDE = 64; this is shown in Figure~\ref{fig:systematics_m10}.

For the \unWISE scanning law, using the $\sim$10.6 million \unWISE sources described above, we take the mean number of single-exposure images in the coadd in $W1$ for the sources in each NSIDE = 64 healpixel.
This is shown in Figure~\ref{fig:systematics_unwisescan}; we can see that the scan is in strips of constant ecliptic latitude, and that there is a significant increase in observations at the ecliptic poles.

To model the selection function we use a Gaussian process, a flexible machine-learning method for regression; for a detailed treatment, see \cite{RasmussenWilliams2006}.
(We first tried a linear model and found that it gave a very poor fit, because there are significant nonlinearities between the systematics and the catalog number density.)
We first scale the data: for the labels (number of \cat sources per pixel) we work in their logarithm, and only fit for the pixels with a nonzero number of sources.
For the \Gaia stellar distribution, \new{the \unWISE source distribution, the \unWISE scan pattern,} and LMC/SMC map templates, we also take the log of the number of quasars per pixel; for the LMC/SMC map, we first replace zeros with a very small value.
For all of the input feature maps, we take the mean-subtracted systematics values.
We assume a Poisson error on the labels (and apply the appropriate log transformation).
For the Gaussian process, we use the \texttt{george} software package \citep{Ambikasaran2016}. 
We use an exponential squared kernel $k$ of the form
\begin{equation}
    k(r^2) = \text{exp}\left(\frac{-r^2}{2}\right)~,
\end{equation}
where $r$ is the distance between points in feature space.
We train the Gaussian process on all of the data, optimizing the parameter vector using the BFGS solver \citep{fletcher_practical_1987}; this includes fitting for the mean of the labels.
We finally evaluate the predicted number of sources in each pixel.
Where there were no \cat sources in the label map, we fix the prediction to zero.

To convert this to a selection function in terms of the \new{relative completeness}, we first identify ``clean'' pixels in the map having low dust extinction ($A_V < 0.03$ mag), low star counts ($N_\mathrm{stars} < 15$), \new{low \unWISE source counts ($< 150$)}, no stars or \unWISE sources in the LMC or SMC, and high $M_{10}$ ($M_{10} > 21$ mag) and \unWISE coadds ($>150$); this results in \new{479 pixels}.
We take the mean predicted number of quasars in these clean pixels, \new{and add two times the standard deviation in these pixels to encompass the scatter.
We then normalize the predicted source numbers by this value, which ensures that all final values end up being less than 1 for clarity.}
The result is a selection function map in terms of the \new{\emph{relative} probability} of \new{a source at a given location being included} in the catalog.
\new{We emphasize that this is relative; we have not normalized it to an absolute probability so as not to make the selection function map extremely sensitive to the maximum value.}
\new{We also note that this} fit must be redone for each version of the catalog because it depends on the particular number density and distribution of sources.

There will be a dependence of the selection function on the $G$-band magnitude, as well as other quantities such as redshift.
While we do not include these in our modeling or fiducial selection function map, we do release selection functions for a redshift split version of the catalog, using two redshift bins, which is important for certain cosmological analyses.
The code to generate the selection function for any input catalog is also provided so that users can construct maps that meet their needs.

\section{Catalog: Results and Verification}
\label{sec:catalog}

\subsection{Properties of the catalog}
\label{sec:properties}

\begin{figure*}
    \centering
    \subfloat[\label{fig:gcatlo_2d}]{\includegraphics[width=0.7\textwidth]{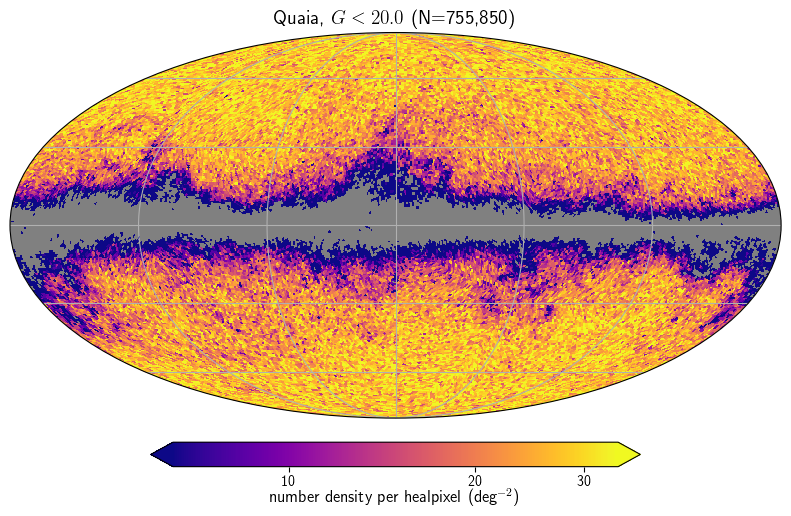}}
    
    \subfloat[\label{fig:gcathi_2d}]{\includegraphics[width=0.7\textwidth]{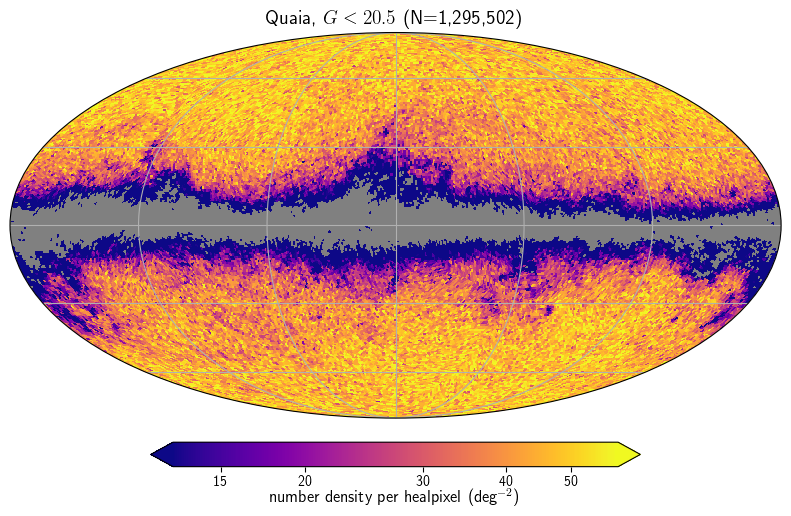}}

    \caption{Sky distribution of the \cat quasar catalog, in Galactic coordinates and displayed using a Mollweide projection. Panel (a) shows sources with $G<\Glo$, the cleaner version with more reliable redshifts, and (b) shows the catalog down to its magnitude limit of $G<\Ghi$.}
    \label{fig:gcat_2d}
\end{figure*}

\cat, the \catalog, consists of \val{N_gcatlo} (\val{N_gcathi}) quasar candidates with $G<\Glo$ $(\Ghi)$.
The sky distribution of \cat for each of these magnitude limits is shown in Figure~\ref{fig:gcat_2d}.
The catalog covers the full sky, besides the Galactic plane, including the southern sky---most of which is not well covered by other surveys (discussed further in \S\ref{sec:comparison}). 
The sky distribution is remarkably uniform, and the nonuniform imprints visually follow the selection effects that we incorporated into our selection function map, most notably the dust distribution (Figure~\ref{fig:systematics_dust}). 
\cat also does not show an obvious overdensity around the LMC and SMC (as the \Gaiapurer sample does) because we have removed these with our decontamination procedure.
In fact, there is now a slight underdensity of sources near the LMC; this makes sense because some quasars in that sky region are obscured by dust and confusion in the LMC, though it is possible we have also somewhat overcorrected for this and removed some true quasars. 

The dearth of quasars in the Galactic plane is due largely to dust extinction and stellar crowding, as well as the fact that the \SDSS training set quasars (for both the original \Gaia DR3 quasar candidates sample and our decontamination procedure) are not representative of quasars in this dust-reddened region. 
If we exclude the regions with very high extinction $A_V>\val{Avhi}$ mag, the quasars nearly uniformly cover the remaining sky area, which comprises \val{area_below_Avhi} ($f_\text{sky}=\val{fsky_below_Avhi}$).
Based on this area we can also compute the effective volume $V_\mathrm{eff}$ covered by the quasars, which depends on the number density as a function of redshift and the power spectrum value $P(k)$, integrated over the physical volume.
We assume a $P(k)$ of $4 \times 10^4 \: (\Mpch)^3$, based on the value for the eBOSS \new{clustering catalog of quasars} at around $k \sim 0.01$ \citep{mueller_clustering_2021}.
This gives an effective volume of \val{volume_effective_gcathi_below_Avhi} (\val{volume_effective_gcatlo_below_Avhi}) for the $G<\Ghi$ ($G<\Glo$) sample.

\begin{figure*}
    \includegraphics[width=\textwidth]{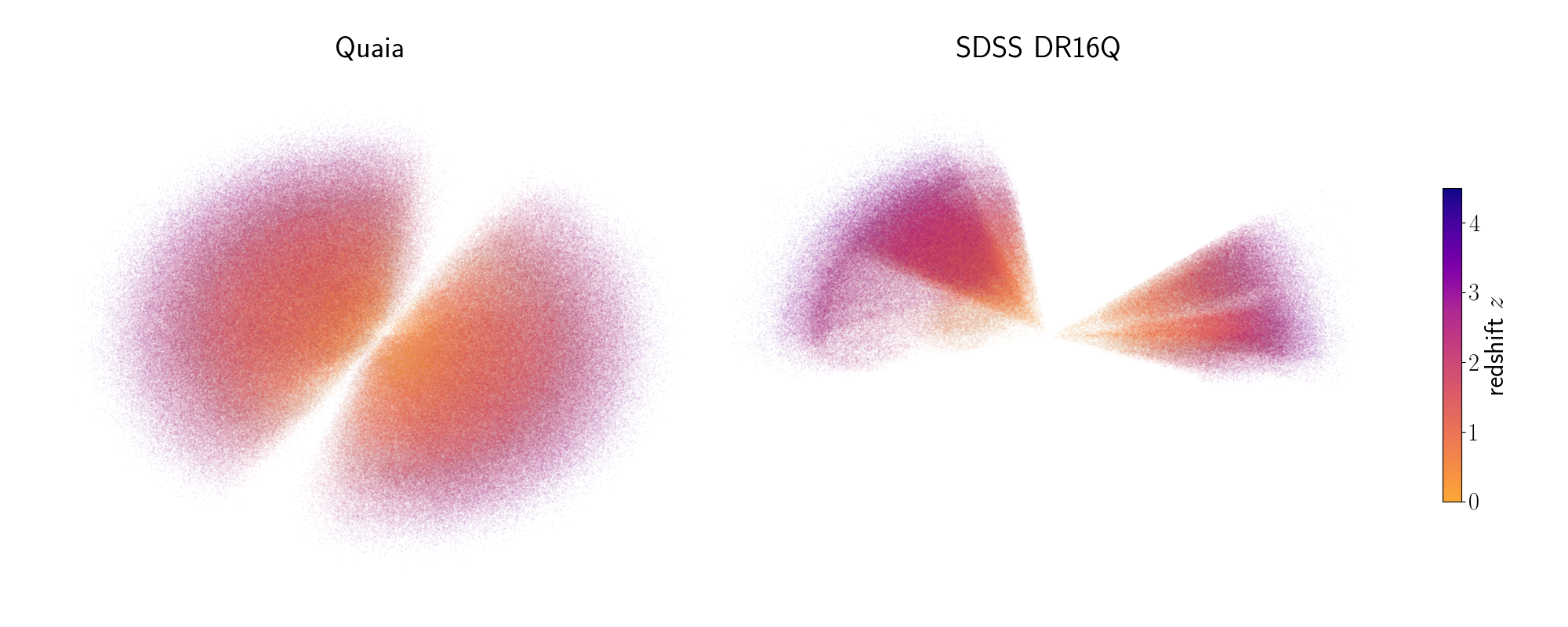}
    \caption{\emph{Left:} a projection of the 3D map of the full \cat catalog ($G<\Ghi$). \emph{Right:} the same projection for the quasars in \SDSS DR16Q\new{, the largest spectroscopic quasar catalog (note that it is a superset of \SDSS quasars from multiple campaigns and as such is not intended to be uniform)}. The color bar shows the redshifts of the quasars ($\zquaia$ for \cat, $\zsdss$ for \SDSS), which have been converted to distances with a fiducial cosmology. \cat spans a \new{significantly larger volume than the \SDSS sample}. 
    A rotating animation of this image is available in the online journal, and at the link in arXiv comment field.
    }
    \label{fig:3d}
\end{figure*}

We show a 3D map of the \cat catalog in Figure~\ref{fig:3d}, using our $\zquaia$ redshift estimates converted to spatial coordinates with a fiducial Planck cosmology.
We also show a 3D map \new{of the full} \SDSS quasar sample for comparison; \cat spans a much larger volume than \SDSS.
\new{We note that} for \SDSS large-scale structure analyses, the eBOSS \new{quasar clustering catalog} is used, which \new{contains fewer sources than the full \SDSS catalog as it} spans only the intermediate (UV-excess) redshift range and is designed to be uniform across the sky (described in more detail in \S\ref{sec:comparison}).

\begin{figure*}
    \centering
    \includegraphics[width=0.9\textwidth]{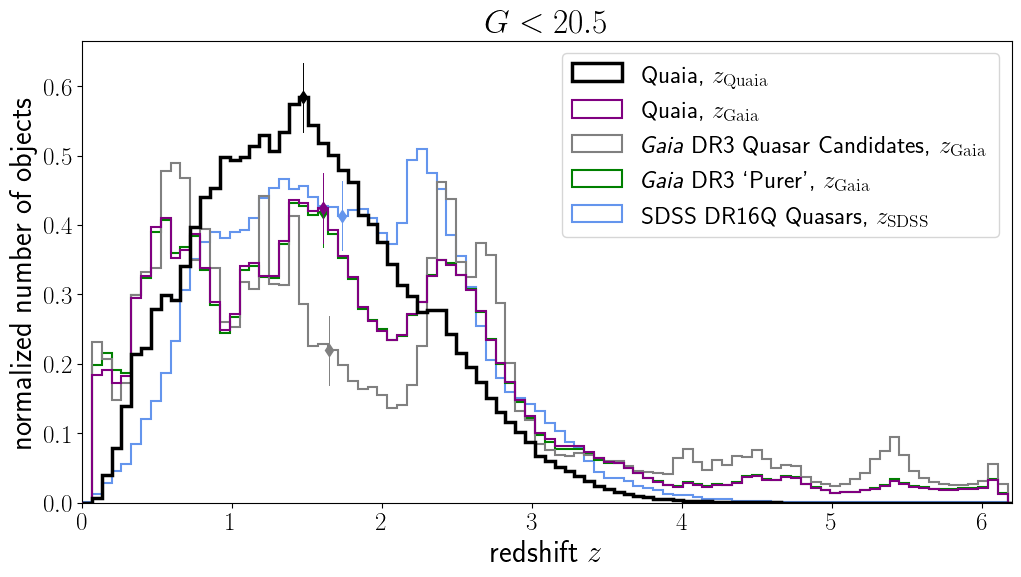}
    \caption{Redshift distribution of \cat for our spectrophotometric redshift estimates $\zquaia$ (black), normalized to the total number of objects. For comparison, we also show the normalized distributions of \new{other samples, cut to the $G<\Ghi$ limiting magnitude of \cat where relevant}: the \Gaia redshift estimates $\zgaia$ for the same \cat sources (purple); $\zgaia$ for the sources in the full \Gaia quasar candidate sample \new{with $G<\Ghi$} (gray); $\zgaia$ for the \Gaiapurer subsample \new{with $G<\Ghi$} (green); and the \SDSS redshifts $\zsdss$ for the \SDSS DR16Q quasar sample \new{that have \Gaia crossmatches, with $G<\Ghi$} (blue). The median redshift of each distribution is shown by the diamond and vertical line in the respective color.} 
    \label{fig:z_dists}
\end{figure*}

We show the redshift distribution of \cat in Figure~\ref{fig:z_dists}.
The distribution of our \Gaia--\unWISE--\SDSS spectrophotometric redshift estimates, $\zquaia$, for the \new{full $G<\Ghi$ catalog} is shown in black.
We compare this to other samples, cut to the same $G$ limit \new{where relevant}: the \Gaia redshifts $\zgaia$ for the same sample; $\zgaia$ for sources in the \new{full} \Gaia quasar candidates sample \new{with $G<\Ghi$} (that have redshift  estimates); $\zgaia$ for sources in the \Gaiapurer sample \new{with $G<\Ghi$} (that have redshift  estimates); and $\zsdss$ for the \SDSS DR16Q sources \new{that have \Gaia crossmatches, with $G<\Ghi$}.
We see that the \cat SPZ redshifts have a smoother distribution than the others, with a clear peak around $z=1.5$; the median value is \val{z_med_gcathi}.
These SPZ estimates have also greatly reduced the high-$z$ tail present in the \Gaia redshifts.
There are still a significant amount of intermediate-$z$ objects; \new{$\val{p_above_zintermediate_gcathi}\%$ ($N=\val{N_above_zintermediate_gcathi}$) of the sources in the full $G<\Ghi$ \cat catalog have $z>\val{zintermediate}$} (for the $G<\Glo$ catalog, this is also $\val{p_above_zintermediate_gcatlo}\%$ ($N=\val{N_above_zintermediate_gcatlo}$) of sources).
We note that the $\zgaia$ redshift distribution for the \Gaiapurer sample is very similar to those same redshift estimates for \cat; this is partially because a very high fraction of the objects in \cat are also in the larger \Gaiapurer sample (see Figure~\ref{fig:frac_matrix}).

We see a slight bump in the $\zquaia$ distribution around $z\sim2.3$, the same location as the peak in the \SDSS DR16Q quasar distribution.
In the \SDSS distribution this feature is most prominent in the \SDSS-III campaign quasars (see Figure 6 of \citealt{lyke_sloan_2020}), which targeted higher-redshift sources.
To check the robustness of our redshift estimation, we reconstruct the sample and retrain the redshifts using the eBOSS \new{quasar clustering catalog} \citep{ross_completed_2020}.
This is the sample used for large-scale structure clustering analyses (e.g. \citealt{mueller_clustering_2021, rezaie_primordial_2021}), which has a smooth redshift distribution peaked around $z=1.5$.
It does still have a slight step around $z\sim2.3$. 
We find that the $\zquaia$ redshift distribution does not change significantly when trained on this sample, and that the feature at $z\sim2.3$ remains.
We hypothesize that this feature is thus a real feature of \Gaia-selected quasars, rather than an imprint from the training set, likely related to details of the optical color selection around that redshift.
We also find that compared to the full \SDSS-trained sample, the sample \new{trained on the eBOSS quasar clustering catalog} produces a redshift distribution that is less smooth at low redshifts, possibly because of the lower number of low-$z$ eBOSS quasars; similarly, the high-$z$ tail is shorter.
For these reasons, we choose to use the full \SDSS sample (as described in \S\ref{sec:data_sdss_quasars}) for the spectroscopic quasar training sample for our fiducial \cat catalog, but confirm that the redshift distribution (and the source selection) is broadly robust to this choice.

\begin{figure}
    \centering
    \includegraphics[width=\columnwidth]{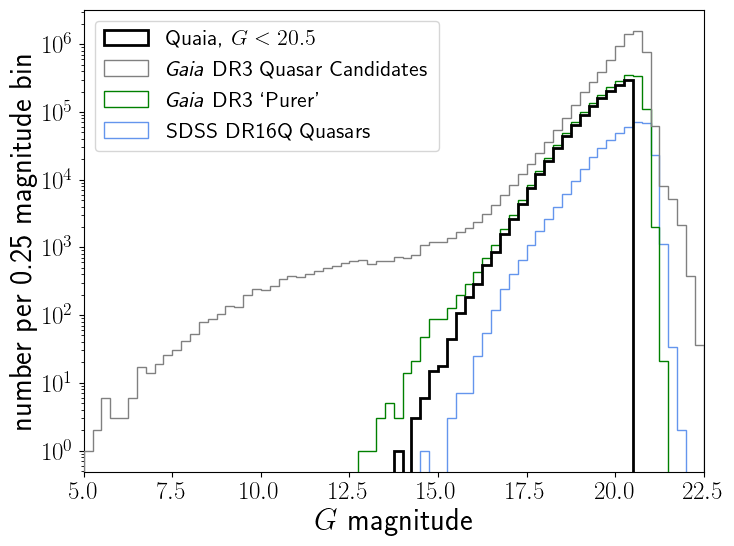}
    \caption{Distribution of $G$ magnitudes of \cat (black), compared to the full \Gaia candidates sample (gray), the \Gaiapurer sample (green), and the \SDSS DR16Q quasar sample (blue).}
    \label{fig:G_dist}
\end{figure}

We show the $G$-band magnitude distribution of \cat in Figure~\ref{fig:G_dist}, in comparison to the other \Gaia and \SDSS quasar samples described above.
We see that our catalog (as well as the \Gaiapurer sample) has removed all of the sources with excessively bright (for quasars) magnitudes $G<12.5$ that are present in the full \Gaia sample, as well as many sources with $12.5<G<16$.
For the \Gaia DR3 and \SDSS samples, the number of quasars drops off sharply after $G\sim20.75$; to avoid the complicated selection effects at these depths, we limit our catalog to $G<\Ghi$ as shown.
We also note that the \SDSS DR16 quasars do not extend as bright as \cat, and this extrapolation past the training set could bias the results in this regime, though in practice this affects very few sources.

\new{We note that some of the \cat sources may technically be considered lower-luminosity AGNs, or Seyfert-like galaxies, rather than quasars.
We estimate the fraction of these sources using the criterion of \cite{schneider_sloan_2010}: sources are considered true quasars if they have \SDSS $i$-band luminosity $M_i$ brighter than $M_i = -22.0$.
To estimate the $i$-band magnitude for our \Gaia sources, we compute the median $G - i$ color for the subset of Quaia sources with \SDSS crossmatches, where $G$ is the \Gaia $G$ band, and then subtract this value from the $G$-band magnitudes to obtain an effective $i$-band magnitude for all \cat sources.
We convert these to absolute magnitudes $M_i$ assuming a flat $\Lambda$CDM cosmology with $H_0 = 70$ km s$^{-1}$ Mpc$^{-1}$, $\Omega_m = 0.3$, and $\Omega_\Lambda = 0.7$, following \cite{schneider_sloan_2010}, and assuming a value of dust reddening of $A_v/E(B-V) = 1.698$ corresponding to the \SDSS $i$ band and $R_v=3.1$.
We find that a small fraction, $\val{p_seyferts_Mi-22}\%$, of \cat sources have effective $M_i < -22.0$ and thus do not meet this standard luminosity criterion for being true quasars.
This distinction may be important for certain studies, though may not be relevant for others, and should be kept in mind for analyses of \cat.}

\subsection{Selection function model}
\label{sec:selfunc}

\begin{figure}
    \centering
    \subfloat[\label{fig:selection_function_map}]{\includegraphics[width=\columnwidth]{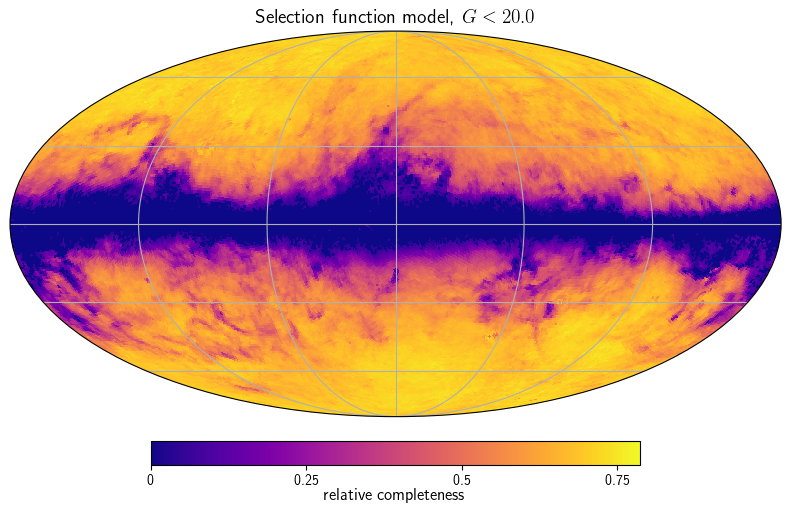}}
    \vspace{0.1ex}
    \subfloat[\label{fig:selection_function_residuals}]{\includegraphics[width=\columnwidth]{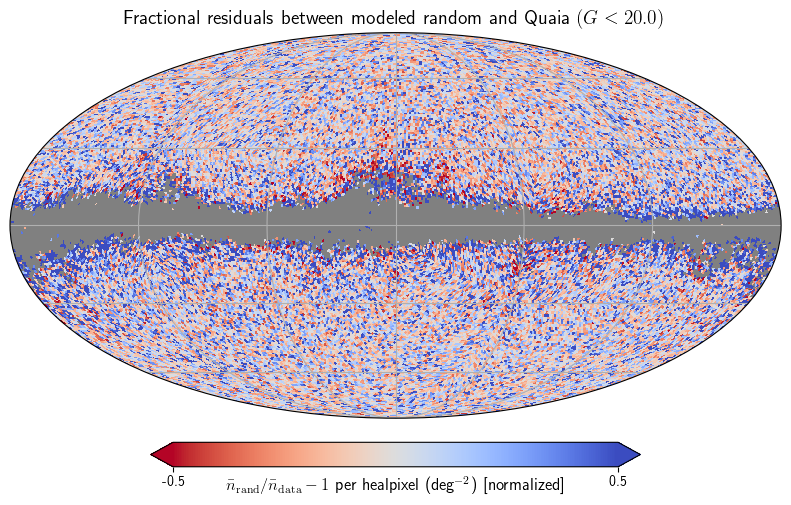}}

    \caption{(a) The selection function map for the $G<\Glo$ subset of \cat, based on a Gaussian process model of the dust, stellar distribution, and $M_{10}$. (b) The fractional residuals between a random catalog downsampled by the modeled selection function and the true \cat $G<\Glo$ catalog.}
    \label{fig:selection_function}
\end{figure}

We show the results of our selection function modeling (\S\ref{sec:selfunc_methods}) for the $G<\Glo$ catalog in Figure~\ref{fig:selection_function}.
The selection function map is shown in Figure~\ref{fig:selection_function_map} \new{, where the values are the \emph{relative} completeness; note that these should not be interpreted as a probability, and users may choose to normalize these values in different ways.}
The relationship \new{of the selection function model} to the dust and \new{source density} maps is clear visually.
In Figure~\ref{fig:selection_function_residuals}, we show the fractional residuals between a random catalog downsampled by this selection function and the true quasar catalog.
The residuals generally look like homogeneous noise, indicating a good fit; the root mean squared fractional error is \val{rmse_fractional_residuals_Glo}.

Around the edges of the Galactic plane the \new{residuals show a slight bias to positive values} (meaning the completeness there was predicted to be higher than it actually is); \new{in the region around zero Galactic longitude just above the Galactic plane, the residuals are slightly biased to negative values (meaning the completeness there was predicted to be lower than it is).
These discrepancies indicate} that our templates are not fully capturing the selection effects \new{in these regions.}
\new{As these are largely limited to the region around the Galactic plane, the} issue could be circumvented by applying a latitude cut for sensitive analyses.
The underdensity around the LMC is well modeled by the selection function, with no clear residual in that region.
The selection function map for the $G<\Ghi$ catalog (not shown) \new{is similar with some moderate differences}, and is also provided as a data product.

\new{The selection function may change more significantly for different subsets of the catalog, such as redshift bins. 
The selection function should be re-fit for a given sample to be analyzed;} we provide code to fit the selection function for any other subset of the catalog.
\new{We note that depending on the subsample, certain regions may be more poorly modeled, and in particular, the regions around the LMC and SMC; users should check the residuals and may choose to mask the regions around the LMC and SMC to be more conservative.}

\subsection{Comparison to existing quasar catalogs}
\label{sec:comparison}
\begin{figure*}
    \centering
    \subfloat[\label{fig:2d_gpurer_Ghi}]{\includegraphics[width=0.45\textwidth]{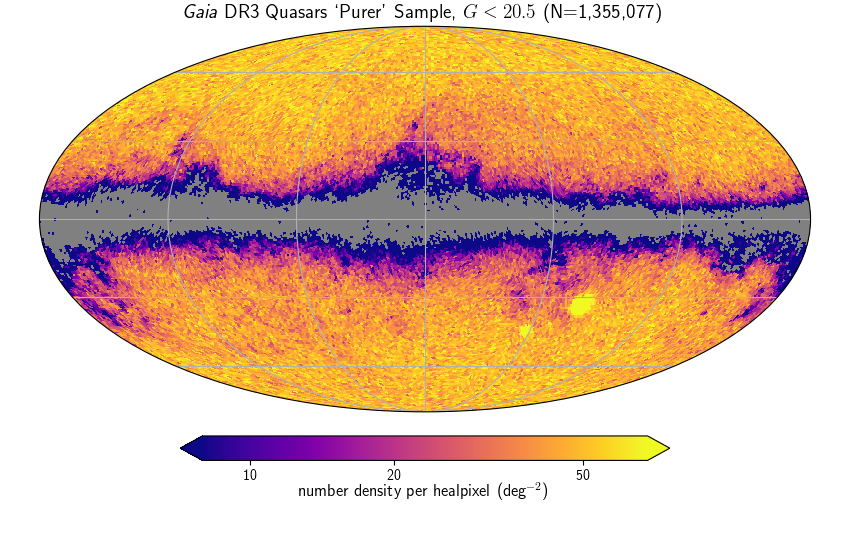}}
    \hspace{5ex}
    \subfloat[\label{fig:2d_wiseps}]{\includegraphics[width=0.45\textwidth]{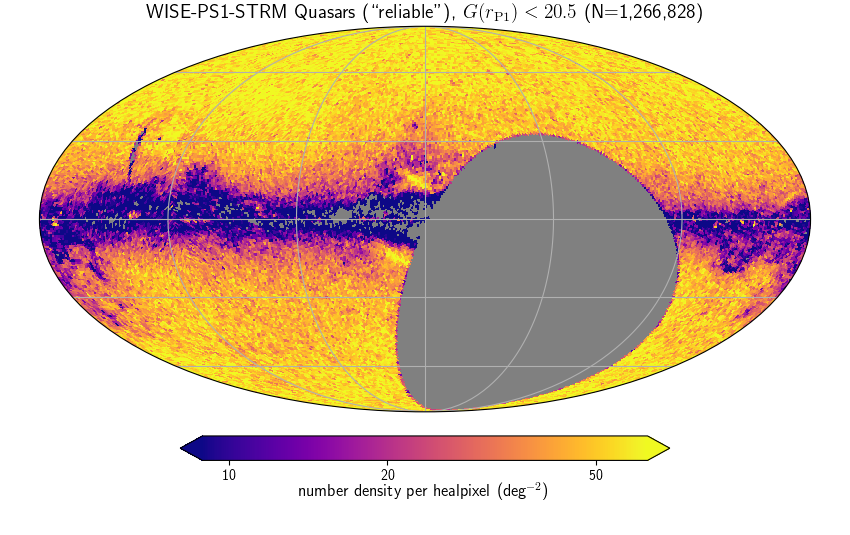}} \\

    \subfloat[\label{fig:2d_sdss}]{\includegraphics[width=0.45\textwidth]{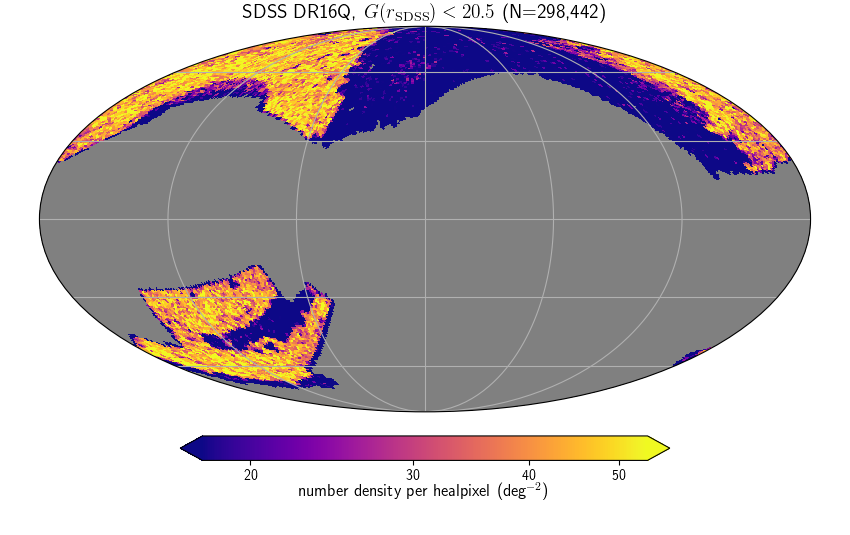}}
    \hspace{5ex}
    \subfloat[\label{fig:2d_eboss}]{\includegraphics[width=0.45\textwidth]{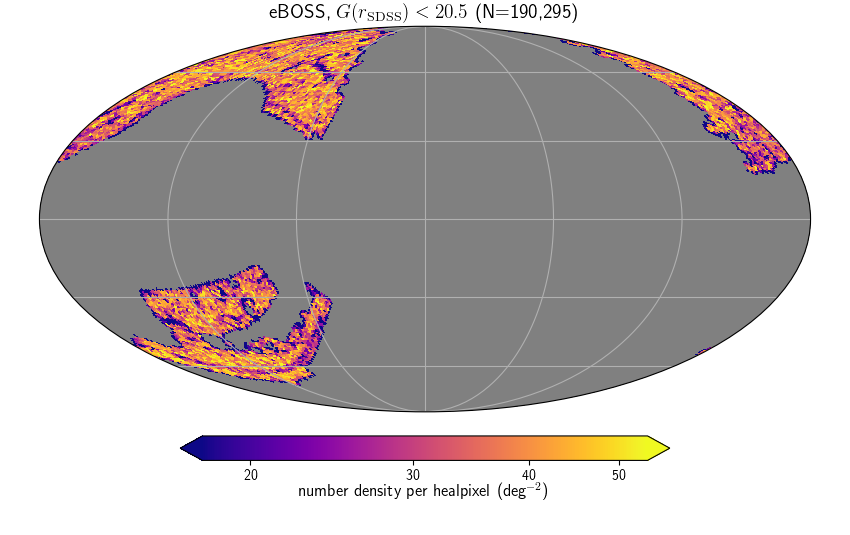}} \\

    \subfloat[\label{fig:2d_milliquas}]{\includegraphics[width=0.45\textwidth]{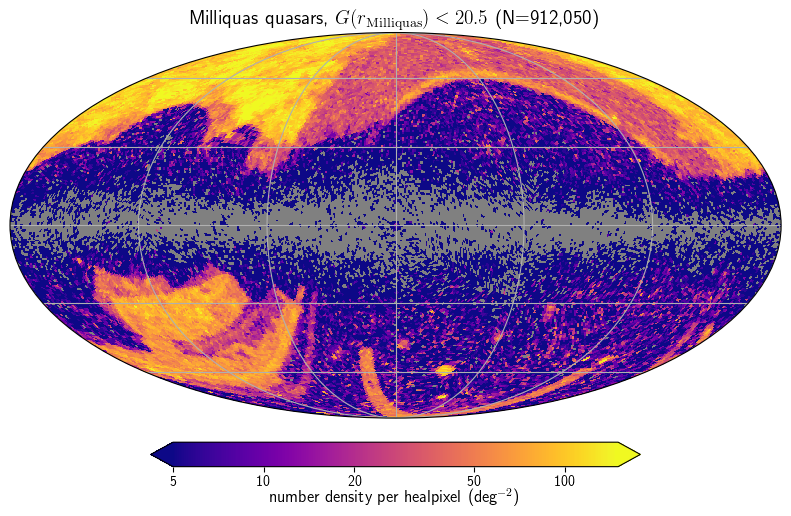}}

    \caption{Other current quasar catalogs for comparison with \cat. All are shown for sources with $G<\protect\Ghi$ or the equivalent converted from another band, in Galactic coordinates and displayed using a Mollweide projection. The catalogs are (a) the \Gaiapurer sample, (b) the WISE-PS1-STRM catalog, \new{(c) the \SDSS DR16Q catalog, (d)} the eBOSS \new{quasar clustering catalog}, and \new{(e)} the Milliquas catalog. Note that the color bars have different scales in each panel.}
    \label{fig:2d_comp}
\end{figure*}

We compare \cat to other existing quasar catalogs: 
Projections of these catalogs are shown in Figure~\ref{fig:2d_comp}. 
We show the \Gaiapurer sample (Figure~\ref{fig:2d_gpurer_Ghi}); a crossmatched catalog of WISE and Pan-STARRS (WISE-PS1), a current leading large-area photometric redshift quasar sample (Figure~\ref{fig:2d_wiseps}); \new{the \SDSS DR16Q catalog, the current best spectroscopic sample of quasars (Figure~\ref{fig:2d_sdss})}; \new{the eBOSS quasar clustering catalog, the subsample of \SDSS DR16Q intended for clustering analyses (Figure~\ref{fig:2d_eboss})}; and  Milliquas, a meta-catalog compiling confirmed quasars from the literature (Figure~\ref{fig:2d_milliquas}).

The \Gaiapurer sample is described in \S\ref{sec:data_gaia}; here we include only sources with QSOC redshift estimates ($\zgaia$).
The WISE-PS1 sample was constructed by \cite{beck_wise-ps1-strm_2022}, based on the Source Types and Redshifts with Machine learning (STRM) algorithm by \cite{beck_ps1-strm_2020}.
The quasar catalog with updated photometric redshifts is presented by \cite{kunsagi-mate_photometric_2022}; here we include only those quasars with redshifts labeled ``reliable'', which is $\val{p_reliable_wiseps}\%$ of the sample.
\new{The \SDSS DR16Q quasar catalog is the one described in \S\ref{sec:data_sdss_quasars}, from \cite{lyke_sloan_2020}, which compiles sources from eBOSS as well as previous \SDSS campaigns (and is intended as a superset of \SDSS quasars rather than a uniform sample).}
The eBOSS \new{quasar clustering} catalog is detailed in \cite{ross_completed_2020}; \new{it is a subsample of \SDSS DR16Q selected for large-scale structure clustering analyses, and as such is much more homogeneous than the full catalog.} 
\new{For the eBOSS clustering catalog, we} have included both eBOSS and legacy \SDSS quasars (\texttt{IMATCH}=1 or 2) and applied the clustering cuts of requiring sectors to have $>0.5$ completeness (\texttt{COMP\_BOSS}) and redshift success rate (\texttt{sector\_SSR}); we have additionally removed sources with \texttt{ZWARNING}!=0.
The Milliquas catalog was compiled by \cite{flesch_million_2021}; a significant portion of the sources are from \SDSS and \textsl{AllWISE}.
For each of these samples, we have shown quasars brighter than a limiting magnitude of $G\sim\Ghi$; for the non-\Gaia catalogs we convert to $G$ from the survey's $r$-band magnitude using the conversion in equation (2) of \cite{proft_exploration_2015}, which is based on the \SDSS $r'$ band.
While this should give a reasonable estimate for the \SDSS sample (using $r_\text{SDSS}$) and the WISE-PS1 sample (using $r_\text{PS1}$ which is very similar to $r_\text{SDSS}$), it may not be as reliable for Milliquas which catalogs ``red'' magnitudes from various sources, as well as for sources with $z>3$ which were not included in the \cite{proft_exploration_2015} fit.

\begin{table*}[h]
    \centering
    \caption{Comparison between \cat and Other Existing Quasar Catalogs, \new{Detailed in the Text}. \new{We show the quantities for the full catalogs (for sources with reliable redshifts) as well as the catalogs limited to} $G<\protect\Ghi$ or the rough equivalent converted from another band. \new{For all quantities and catalogs shown,} we exclude areas with high dust extinction ($A_V > \protect\val{Avhi}$ mag)\new{; this excludes $\sim$5\% of sources for \cat and \Gaiapurer, $\sim$18\% of the full WISE-PS1 sample, and a negligible number of sources for \SDSS DR16Q and the eBOSS clustering catalog.} \new{We note that the \SDSS DR16Q catalog is a superset of quasars from many \SDSS campaigns and is not intended to be uniform, which should be considered in particular for the sky fraction and spanning volume quantities.} We show the number of sources $N$, the fraction of sky area covered $f_\mathrm{sky}$, the mean number density per square degree $\bar{n}$, the spanning volume between $0.8<z<2.2$ $V_\mathrm{span}$, the effective volume $V_\mathrm{eff}$, the median redshift $z_\mathrm{med}$, and the fraction of objects with $|\delta z| \equiv |\dz| < 0.01$ and $<0.1$ (where applicable).}
    \begin{tabular}{|l|l|l|p{2.5em}|p{4.5em}|p{4.5em}|l|p{7em}|p{7em}|}
\hline
 & $N$ & $f_\text{sky}$ & $\bar{n}, \newline \text{deg}^{-2}$ & $V_\text{span}, \newline (h^{-1}\,\text{Gpc})^3$ & $V_\text{eff}, \newline (h^{-1}\,\text{Gpc})^3$ & $z_\text{med}$ & $f(|\delta z| <0.01)$ & $f(|\delta z| <0.1)$ \\
\hhline{|=|=|=|=|=|=|=|=|=|}
\textbf{Quaia} & \textbf{1,234,715} & \textbf{0.73} & \textbf{40.78} & \textbf{143.78} & \textbf{7.08} & \textbf{1.48} & \textbf{0.63} & \textbf{0.84} \\
\hline
\emph{Gaia} Purer & 1,647,311 & 0.73 & 54.42 & 143.76 & 9.24 & 1.63 & 0.53 & 0.62 \\
\hspace{1.5em}$G<20.5$ & 1,286,788 & 0.73 & 42.51 & 143.76 & 6.50 & 1.61 & 0.62 & 0.70 \\
\hline
WISE-PS1 & 2,386,121 & 0.56 & 103.89 & 109.08 & 20.88 & 1.38 & 0.11 & 0.71 \\
\hspace{1.5em}$G_\mathrm{eff}<20.5$ & 1,130,925 & 0.56 & 49.25 & 109.06 & 7.32 & 1.41 & 0.12 & 0.76 \\
\hline
SDSS DR16Q & 637,371 & 0.26 & 60.18 & 50.30 & 4.16 & 1.77 & $\sim$1 & $\sim$1 \\
\hspace{1.5em}$G_\mathrm{eff}<20.5$ & 297,940 & 0.26 & 28.17 & 50.23 & 1.18 & 1.67 & $\sim$1 & $\sim$1 \\
\hline
eBOSS Clustering & 409,286 & 0.14 & 72.52 & 26.80 & 3.21 & 1.60 & $\sim$1 & $\sim$1 \\
\hspace{1.5em}$G_\mathrm{eff}<20.5$ & 190,263 & 0.14 & 33.96 & 26.61 & 1.01 & 1.49 & $\sim$1 & $\sim$1 \\
\hline
\end{tabular}
    \label{tab:comparison}
\end{table*}

A summary of the catalogs is shown in Table~\ref{tab:comparison}, for \new{the full catalogs (limited to sources with reliable redshifts) as well as} the $G_\mathrm{eff}<\Ghi$ subsamples.
We exclude Milliquas from this comparison given its very heterogeneous nature \new{; we do include \SDSS DR16Q, though it is also not intended to be uniform, to show the comparison of \cat to this large spectroscopic catalog of quasars}.
For these quantifications, we exclude areas that have $A_V > \val{Avhi}$ mag, as well as healpixels with no quasars.
For the sky fraction $f_\mathrm{sky}$, we see that \cat and \Gaiapurer are limited only by the dusty regions, and cover over 30\% more area than WISE-PS1 (which is limited by Pan-STARRS), \new{nearly $3\times$ that of \SDSS DR16Q, and over $5\times$ that of the eBOSS quasar clustering catalog}.
Compared to the \Gaiapurer sample, \cat has a slightly smaller number of sources, but due to its redshift distribution gives a slightly higher effective volume.
The on-sky number density is similar for all of the catalogs \new{when limiting them to similar magnitudes}, with WISE-PS1 slightly higher because it has a similar number of objects to the \Gaia catalogs but over a smaller area, \new{and \SDSS DR16Q and the eBOSS clustering catalog slightly lower}.
\new{When including faint sources for the catalogs, WISE-PS1 has $2.5\times$ the on-sky number density as \cat, and \SDSS DR16Q and the eBOSS clustering catalog have $1.5-2\times$.}

For the volume comparison, we compute two different volumes. 
The first is a simple ``spanning'' volume, $V_\mathrm{span}$, which is just the comoving volume in the sky area of the survey (as given by $f_\mathrm{sky}$ of the full sky area) in a redshift range $0.8<z<2.2$, a typical redshift range for clustering analyses (taken from the range of the eBOSS \new{quasar clustering catalog}).
Thus it compares in the same way as the survey areas, but gives an idea of the physical volume the catalogs span.
The second is the effective volume, described in \S\ref{sec:properties}; we use that same $P(k) = 4 \times 10^4 \: (\Mpch)^3$ for the volume calculation for all catalogs.
We see that the effective volume \new{of WISE-PS1 is much larger (nearly $3\times$) than that of \cat as a result of its larger number of sources, though when considering samples with the same limiting magnitude, WISE-PS1 and \cat have comparable effective volumes.} 
\new{The effective volume of \cat is nearly twice as large as that of \SDSS DR16Q, and $6\times$ for the magnitude-limited sample; compared to \new{the eBOSS quasar clustering catalog}, the effective volume of \cat is over twice as large, and $7\times$ for the magnitude-limited sample.}

The catalogs all have a similar median redshift, of around \new{$1.4 < z < 1.7$, extending to 1.77 for \SDSS DR16Q when including faint sources}.
However, they have significantly different redshift precision; in Table~\ref{tab:comparison} we show \new{outlier fractions estimated from} comparisons to spectroscopic redshifts.
We see that both of the \Gaia catalogs have a similar fraction of high-precision redshifts ($|\dz| < 0.01$), but \cat has a much higher fraction of redshifts that are not strong outliers ($|\dz| < 0.1$) compared to \Gaiapurer.
WISE-PS1 falls between \new{\cat and \Gaiapurer} in terms of strong outliers, but has an extremely low fraction of high-precision redshifts as it is a photometric survey.
\new{We note that for both \Gaiapurer and WISE-PS1, the redshift precision is significantly lower when considering the full catalog compared to samples limited to $G_\mathrm{eff}<20.5$ like \cat; we show both for a fair comparison.}
The \new{\SDSS DR16Q catalog and the eBOSS quasar clustering catalog have} spectroscopic redshifts, so these are almost all very high precision; \cite{lyke_sloan_2020} estimated from a visual inspection that less than 1\% of the \SDSS DR16Q redshifts are outliers with $\Delta v > 3000\,\text{km s}^{-1}$ ($|\Dz| > 0.01$), independent of redshift; note that this is a slightly different sample than the eBOSS \new{clustering catalog}, but we can expect it to be similar.
The \SDSS DR16Q quasar sample has typical statistical redshift errors of $|\Dz| \sim 0.001$.

To give more of an idea of the redshift precision of \cat, we compare it to existing all-sky photometric galaxy catalogs.
A common statistic to summarize photometric redshift uncertainty robust to outliers is the SMAD, scaled median absolute deviation, defined as $1.4826 \times \text{med}(|\Dz - \text{med}(\Dz)|)$, where $\Dz = z_\text{phot} - z_\text{spec}$ (the scaling factor adjusts the MAD such that SMAD is approximately equal to the standard deviation for normalized data).
The SMAD of the full \cat catalog ($G<\Ghi$) is SMAD($\Dz$) = \val{smad_Dz_Ghi}, and the normalized SMAD of the redshift errors with the $(1+z)$ factor divided out is SMAD($\dz$) = \val{smad_dz_Ghi}.
For comparison, the WISE $\times$ SuperCOSMOS catalog of 20 million galaxies with $z_\text{med} = 0.2$ \citep{bilicki_wise_2016} has an SMAD($\Dz$) of $\sim 0.04$ and an SMAD($\dz$) of $\sim 0.035$.
The Two Micron All Sky Survey Photometric Redshift (2MPZ) catalog has around 1 million galaxies with a similar median redshift \citep{bilicki_2mass_2013}, which have an SMAD($\Dz$) of $\sim 0.015$. 
\cat thus falls in between these common photometric galaxy samples in terms of overall redshift precision; however, we note that it is difficult to capture the redshift error of \cat in a single statistic, given both its large number of highly precise redshifts and non-insignificant number of outliers.

We also note that the ongoing DESI survey \new{\citep{Aghamousa2016, desi_collaboration_validation_2023}} will observe a high density of quasars over a large sky area \new{\citep{chaussidon_target_2023}}, which will be competitive with and complementary to \cat.

\subsection{Catalog format}
\label{sec:format}

\begin{table*}
    \caption{Format and Column Descriptions of \cat, Published as a FITS Data File \citep{wells_fits_1981}. For the example entry, we show the first catalog row.}
    \centering
    \begin{tabular}{|l|l|l|p{24em}|l|}
\hline
column name & symbol & units & description & example entry value \\
\hline
\texttt{source\_id} &  &  & \emph{Gaia} DR3 source identifier & 6459630980096 \\
\texttt{unwise\_objid} &  &  & unWISE DR1 source identifier & 0453p000o0014479 \\
\texttt{redshift\_quaia} & $z_\mathrm{Quaia}$ &  & spectrophotometric redshift estimate & 0.416867 \\
\texttt{redshift\_quaia\_err} &  &  & $1\sigma$ uncertainty on spectrophotometric redshift estimate & 0.060812 \\
\texttt{ra} &  & deg & barycentric right ascension of the source in ICRS at 2016.0 & 44.910498 \\
\texttt{dec} &  & deg & barycentric declination $\delta$ of the source in ICRS at 2016.0 & 0.189649 \\
\texttt{l} &  & deg & galactic longitude & 176.659434 \\
\texttt{b} &  & deg & galactic latitude & -48.835164 \\
\texttt{phot\_g\_mean\_mag} & $G$ & mag & \emph{Gaia} $G$-band mean magnitude & 20.173105 \\
\texttt{phot\_bp\_mean\_mag} & $BP$ & mag & \emph{Gaia} integrated $BP$ mean magnitude & 20.200150 \\
\texttt{phot\_rp\_mean\_mag} & $RP$ & mag & \emph{Gaia} integrated $RP$ mean magnitude & 18.871586 \\
\texttt{mag\_w1\_vg} & $W1$ & mag & unWISE $W1$ magnitude & 14.774343 \\
\texttt{mag\_w2\_vg} & $W2$ & mag & unWISE $W2$ magnitude & 13.923867 \\
\texttt{pm} & $\mu$ & mas / yr & total proper motion & 0.383797 \\
\texttt{pmra} & $\mu_{\alpha*}$ & mas / yr & proper motion in right ascension $\mu_{\alpha*} \equiv \mu_\alpha \, \mathrm{cos} \, \delta$ of the source in ICRS at 2016.0 & 0.217806 \\
\texttt{pmdec} & $\mu_{\delta}$ & mas / yr & proper motion in declination $\mu_{\delta}$ of the source in ICRS at 2016.0 & -0.316007 \\
\texttt{pmra\_error} & $\sigma_{\mu\alpha*}$ & mas / yr & standard error of proper motion in right ascension direction & 0.679419 \\
\texttt{pmdec\_error} & $\sigma_{\mu\delta}$ & mas / yr & standard error of proper motion in declination direction & 0.608799 \\
\hline
\end{tabular}
    \vspace{2ex}
    \label{tab:catalog}
\end{table*}

The complete \cat catalog contains our decontaminated quasar sample with computed redshift information, relevant \Gaia properties, and crossmatched catalog information.
The complete catalog format with column names, units, column descriptions, and an example entry is shown in Table~\ref{tab:catalog}.
Additional information for the sources can be obtained by joining the catalog with the relevant data source with the associated identifier (\Gaia or \unWISE).
We include only sources with $G<\Ghi$ in the catalog; we also publish a version limited to $G<\Glo$, along with the selection function models fit to each (\S\ref{sec:selfunc}) and ``random'' catalogs generated from the selection functions.
The catalog includes our SPZ redshifts $\zgaia$ along with 1$\sigma$ redshift errors, sky position, \Gaia photometry, \unWISE photometry, and proper motion information.
The catalog is in FITS format \citep{wells_fits_1981}, and units and descriptions are provided for each column.

\subsection{Limitations}
\label{sec:limitations}

While the \cat catalog presents a highly useful quasar sample, it does have various limitations.
We reiterate and discuss the main ones here.

We estimate spectrophotometric redshifts for the quasars, which are generally more accurate than the \Gaia estimates, but are still low precision compared to spectroscopic redshifts. 
The uncertainties on these redshifts should be taken into account for any measurements, and the rate of catastrophic redshift errors (not necessarily captured by the redshift uncertainty) should be considered when thinking about possible uses of the catalog.

The selection function model has multiple potential limitations. 
While it broadly captures the selection effects that affect the quasar sample, it has significantly lower accuracy around the galactic plane; precision measurements may require masking this region.
The regions around the LMC and SMC are also more poorly modeled; users may want to mask this area.
We also note that we are not fitting the healpixels with zero quasars, which may result in a slight bias toward populated regions, and fixes the zero-probability region of the selection function.
Our selection function map depends only on sky position and not other properties such as magnitude or redshift (besides fitting it to the appropriate subsample); a treatment incorporating these dependencies may be important for certain uses.
The gold standard for completeness estimation is data injection and recovery tests.
Unfortunately, the \Gaia instrumentation has black-box elements, such as onboard image segmentation, onboard object detection, and onboard downlink prioritization, that make it impossible to perform end-to-end injection tests, so we rely on a data-driven approach, which may be less robust and more sensitive to modeling choices.
Given this, it is possible that we are overfitting the selection function.
Finally, the selection function depends on the assumption of isotropy, which we know to be broken to some extent by the kinematic dipole \citep{stewart_peculiar_1967, secrest_test_2021}; we will explore and measure this in an upcoming work (see \S\ref{sec:applications}).
Users employing the selection maps or generating their own selection function for some subset of the catalog should take note of these potential issues.

Generally, \cat has relatively low number density (e.g. compared to the \SDSS sample). 
This means that it may not be ideal for certain cosmological measurements, which may be shot noise dominated. 

Finally, we note that this catalog is based on the \Gaia quasar candidates sample, and it will inherit many of the limitations of that sample \citep{gaia_collaboration_gaia_2023}.
We are also limited to the \Gaia-derived properties (e.g. the \Gaia redshifts that are a feature for our estimates).
In upcoming \Gaia data releases, the collaboration will release more BP/RP spectra and we will have the opportunity to work directly from the spectral data to improve the catalog.

\subsection{Potential applications}
\label{sec:applications}

Quasars are highly biased tracers of the cosmic web that trace the matter distribution at higher redshift than galaxies and in the mildly nonlinear regime. 
Given the \cat catalog's sampling of quasars to deep magnitudes and across a large volume, and its reduced systematic contamination allowed by space-based observations, \cat lends itself to large-scale structure analyses, many of which are currently ongoing.

Thanks to its large volume and well-characterized selection function, \cat is perhaps the best current sample for testing homogeneity and isotropy in the Universe \citep{quaia-homogeneity}, and relatedly for measuring the dipole in the quasar distribution \citep{quaia-isotropy}, which recent measurements have consistently found to be in mild tension with the kinematic interpretation in the $\Lambda$CDM model.
\cat's volume also makes it a good sample for a measurement of the matter-radiation equality scale, $k_\text{eq}$ (e.g. \citealt{bahr-kalus_measurement_2023}).

The catalog is particularly well suited for cross-correlations with other all-sky observations of projected tracers of the large-scale structure, which are less sensitive to redshift errors compared to 3D ones. 
Examples of this are the CMB, the CIB, or maps of the thermal Sunyaev--Zel'dovich effect. 
\cite{alonso_constraining_2023} used the cross-correlation between CMB lensing and \cat to constrain the growth of matter fluctuations via the parameter $S8$, \new{achieving competitive constraints as well as showing that \cat can break the degeneracy between $\Omega_m$ and $\sigma_8$}.
\new{An analysis of primordial non-Gaussianity (parameterized by $f_{NL}$) from this cross-correlation with CMB lensing is also underway.}
Analyses of the cross-correlation with CMB temperature to measure the Integrated Sachs--Wolfe effect, and with the CIB to constrain the star formation history at high redshifts (e.g. \citealt{jego_star_2023}), are currently under investigation.
Another measurement enabled by the catalog is the cross-correlation of quasar proper motions with the large-scale structure, which gives a direct estimate of the cosmological quantity $H f \sigma_8$ \citep{duncan_can_2023}.
Additionally, cross-correlations \new{of \cat} with galaxy surveys \new{may allow for measurements of} the baryon acoustic feature \new{\citep{patej_analysing_2018,zarrouk_baryon_2021}} and quasar environments \new{\citep{padmanabhan_real-space_2009,shen_cross-correlation_2013}}. 

\cat is also useful for void studies, including constraining core cosmological parameters with the void size distribution; this investigation is underway \citep{quaia-voids}.
The catalog is additionally relevant to astrophysical analyses of quasar properties, given its large sky coverage and multiband photometry, \new{such as the} role of galaxy interactions on AGN activity.
\cat sources may also be used to study the potential of quasars as standard candles.
Further, \cat provides perhaps the best quasar coverage of the southern sky, which may be important for a variety of applications such as identifying interesting sources there, adding new information to known sources, or calibrating surveys in that sky region.
Finally, while a 3D clustering analysis of \cat may be limited by the catalog's relatively low number density and moderate redshift precision, a careful analysis may yield useful constraints, especially using techniques targeted at wide-field surveys (e.g. \citealt{lanusse_3d_2015}). 

The latter is comparable or better than other state-of-the-art galaxy and quasar samples used in large-scale structure analyses, but not enough to necessarily allow an accurate interpretation.

\section{Summary and Data Access}
\label{sec:summary}

We have constructed a new quasar catalog, \cat, the \catalog, designed for cosmological studies, derived from the \Gaia DR3 quasar candidates sample and using \unWISE photometry to remove contaminants and derive precise redshifts.
Our key contributions and the features of the catalog are as follows:
\begin{itemize}
\setlength\itemsep{0.5ex}
    \item We have decontaminated the \Gaia DR3 quasar candidates sample with proper motion cuts and optimized color cuts based on \Gaia and \unWISE photometry. This reduced the number of known contaminants by \val{factor_reduction_contaminants}, while only excluding $\val{p_sqall_excluded_clean}\%$ of known quasars with respect to the superset of \Gaia quasar candidates (that have \unWISE photometry, \Gaia redshifts, and a $G$-magnitude cut of $G<\Gmax$).  
    \item The catalog extends to a limiting magnitude of $G<\Ghi$ and contains \val{N_gcathi} sources; we also release a brighter, cleaner sample limited to $G<\Glo$, which includes \val{N_gcatlo} sources.
    \item \cat covers the entire sky, only limited by selection effects near the Galactic plane; excluding highly dust-extincted regions ($A_V > \val{Avhi}$ mag), this results in an area of \val{area_below_Avhi} ($f_\mathrm{sky}=\val{fsky_below_Avhi}$).
    \item We have improved the \Gaia redshift estimates using a \knn model trained on these redshifts and \Gaia and \unWISE colors with \SDSS spectroscopic redshift labels, producing spectrophotometric redshifts. The median redshift of the $G<\Glo$ catalog is $z_\mathrm{med}=\val{z_med_gcatlo}$, with $\val{p_acc_zspz_dzhi_Glo}\%$ ($\val{p_acc_zspz_dzlo_Glo}\%$) of redshifts within $|\dz|<\val{dzhi}$ (\val{dzlo}) of \SDSS redshifts. This is a reduction in the number of catastrophic outliers by \val{factor_reduction_outliers_dzhi_Glo} (\val{factor_reduction_outliers_dzmid_Glo}) compared to the \Gaia redshift estimates.
    \item We produced a data-driven model of the selection function, which includes the systematic effects of dust, \new{the source density of the parent surveys \Gaia and \unWISE}, and the scanning laws \new{of the parent surveys}. We used this to generate random catalogs of Poisson-distributed points with \new{similar selection effects to} \cat.
\end{itemize}

The catalog, selection function, and related data products are publicly available at \url{https://doi.org/10.5281/zenodo.10403370}, along with documentation.
The code used to generate this catalog is open source and available at \url{https://github.com/kstoreyf/gaia-quasars-lss}.

\section*{}

The authors are grateful to the members of the \Gaia collaboration, in particular Coryn Bailer-Jones, Morgan Fouesneau, Anthony Brown, Ludovic Delchambre, Tristan Cantat-Gaudin, and Arvind Hughes.
The authors also thank Lyuba Slavcheva-Mihova, Nestor Arsenov, Andras Kovacs, An\v{z}e Slosar, Giulia Piccirilli, Iain Duncan, Abby Williams, Dustin Lang, Mehdi Rezaie, Alex Malz, Lehmann Garrison, and Nathan Secrest for very helpful discussions.
Additionally, the authors thank the members of the Astronomical Data group at the Center for Computational Astrophysics for useful feedback.
\new{The authors are grateful to the anonymous referee, whose feedback has significantly strengthened this work.}
This project was developed in part at the Gaia Fête, hosted by the Flatiron Institute Center for Computational Astrophysics in 2022 June.
This work has made use of data from the European Space Agency mission {\it Gaia} (\url{https://www.cosmos.esa.int/gaia}), processed by the {\it Gaia} Data Processing and Analysis Consortium (DPAC; \url{https://www.cosmos.esa.int/web/gaia/dpac/consortium}).
Funding for the DPAC has been provided by national institutions, in particular, the institutions participating in the {\it Gaia} Multilateral Agreement.
This publication makes use of data products from WISE, which is a joint project of the University of California, Los Angeles, and the Jet Propulsion Laboratory/California Institute of Technology, funded by the National Aeronautics and Space Administration.
We specifically use the unWISE coadds, produced by D. Lang, A.M. Meisner, and D.J. Schlegel.
K.S.F. is supported by the NASA FINESST program under award No. 80NSSC20K1545. 
G.F. acknowledges the support of the European Research Council under the Marie Sk\l{}odowska Curie actions through the Individual Global Fellowship No.~892401 PiCOGAMBAS.
D.A. acknowledges support from the Beecroft Trust, and from the John O'Connor Research Fund, at St. Peter's College, Oxford.
This research made use of computational resources at New York University (NYU); the authors thank the NYU high-performance computing team.

\software{Astropy \citep{the_astropy_collaboration_astropy_2013, the_astropy_collaboration_astropy_2018, the_astropy_collaboration_astropy_2022}, NumPy \citep{harris_array_2020}, IPython \citep{Perez2007}, SciPy \citep{Virtanen2020}, matplotlib \citep{Hunter2007}, healpy \citep{gorski_healpix_2005, zonca_healpy_2019}, george \citep{Ambikasaran2016}}

\bibliography{references,references_extra}

\end{document}